\documentclass[aps,11pt,a4paper]{report}

\usepackage{amsmath,amssymb}
\usepackage[numbers,sort&compress]{natbib}
\usepackage{url}         
\usepackage{bm}
\usepackage{graphicx}
\usepackage{thesis1989}
\usepackage{titlesec}
\titlespacing*{\section}{0pt}{0.5ex}{1.5ex}
\titlespacing*{\section}{0pt}{2.0ex plus 0.5ex minus 0.2ex}{1.0ex plus 0.2ex}
\usepackage[unicode]{hyperref}
\hypersetup{
  colorlinks=true,
  linkcolor=blue,
  citecolor=blue,
  urlcolor=blue
}
\numberwithin{equation}{chapter}
\newcommand{\bodystretch}{1.6}

\begin{document}
\setstretch{1.0}
\setcounter{page}{1}
\begin{titlepage}
\thispagestyle{empty}
\centering

\vspace*{2cm}

University of Adelaide\\
Department of Physics and Mathematical Physics

\vspace{3.5cm}

{\Large\bfseries
POINT CHARGES IN\\[-2pt]
CLASSICAL\\[4pt]
ELECTRODYNAMICS
}

\vspace{3.5cm}

A thesis submitted in partial fulfilment of the requirements for the
degree of Bachelor of Science (Honours) in Physics and Mathematical
Physics at the University of Adelaide

\vspace{3cm}

Jonathan N.\ E.\ Baxter\\
October 27, 1989

\end{titlepage}

\clearpage
\thispagestyle{empty}

\begingroup
\setlength{\parindent}{0pt}
\setlength{\parskip}{0.75em}

\begin{center}
{\Large Author's note (2025)}
\end{center}

This is my honours thesis from 1989, posted to arXiv for archival
purposes.

Chapters 2--5 and the appendices contain the main technical work.

Chapter 6 (\emph{A discussion of renormalization}) is left as written
for historical completeness. It was intended as a speculative
discussion and should not be treated as a reliable modern resolution
of renormalization and self-force issues. It also contains at least
one sign/terminology error regarding time-symmetric
vs.\ time-antisymmetric retarded/advanced field combinations.

Minor transcription and notational corrections, together with corrections to
equations, have been made. Apart from these corrections and this note, the
document is reproduced as written.

\vspace{2cm}

\begin{flushright}
Jonathan Baxter\\
(added December 2025)
\end{flushright}

\endgroup

\clearpage
\setcounter{page}{2}
\tableofcontents
\clearpage
\setstretch{\bodystretch}
\abovedisplayskip=7pt plus 3pt minus 3pt
\belowdisplayskip=7pt plus 3pt minus 3pt
\abovedisplayshortskip=0pt plus 2pt minus 2pt
\belowdisplayshortskip=4pt plus 2pt minus 2pt
\jot=2pt

\begin{center}
{\LARGE\bfseries NOTATION}
\end{center}
\vspace{3cm}

Unless otherwise stated, the underlying space in all of our work is Minkowski
space-time with metric
\[g^{\mu\nu} =
\begin{bmatrix}
-1 & 0 & 0 & 0 \\
0 & 1 & 0 & 0 \\
0 & 0 & 1 & 0 \\
0 & 0 & 0 & 1
\end{bmatrix}.\]
Occasionally we may use $\eta^{\mu\nu}$ in place of $g^{\mu\nu}$.

Four-vectors are denoted by roman characters, $a$, $b$ etc, and their
components are referred to by greek indices ($a^\mu$, $b^\nu$, etc)
where $\mu,\nu \in (0,1,2,3)$. The spatial components ($\mu = 1,2,3$)
we refer to with indices $i$, $j$, $k$ etc. Ordinary three-vectors are
written in bold-face characters, so that if $z$ is a four-vector then
\[
z = (z^0,\mathbf{z}).
\]

We use $,\mu$ to indicate partial differentiation with respect to the
$\mu^{\text{th}}$ coordinate.

A four vector contracted with itself, e.g.\ $a^\mu a_\mu$, is written
as $a^2$, and sometimes we use the abbreviation $a_n = a^\mu n_\mu$
for the contraction of two four-vectors.

\clearpage
\chapter{INTRODUCTION}
\label{chap:1}

When, in 1864, James Clerk Maxwell succeeded in combining into one
theory all of the empirical laws of electromagnetism accumulated by
his ingenious predecessors, the prevailing ideas revolved around
electric currents being a kind of fluid, which interacted with each
other through the intermediary of a field. However in 1881 Helmholtz
and Stoney resurrected an old idea of Gustav Fechner's, that electric
currents might be due to charged particles flowing down a
wire. (Stoney also suggested the name ``electron'' for the negatively
charged particles). The acceptance of atoms as the fundamental
building blocks of matter made the idea of electrons as the ``atoms of
electricity'' seem quite plausible at the time.

In that same year Thomson wrote on the subject of electromagnetic
mass. He showed that the field of a spherical charge has a kinetic
energy
\begin{equation}
E_{\mathrm{elm}} = f\,\frac{e^2}{R c^2}\,\frac{v^2}{2},
\label{eq:ThomsonEnergy}
\end{equation}
where $f$ is a form-factor dependent on the charge distribution, $R$
is the radius of the sphere and $e$ is the total charge. Consequently,
one can identify
\begin{equation}
m_{\mathrm{elm}} = f\,\frac{e^2}{R c^2}
\label{eq:ElectromagneticMass}
\end{equation}
as an ``electromagnetic mass''. Thomson concluded that a spherically charged
particle of mass $m_o$ and charge $e$ would have a total kinetic energy
\begin{equation}
E = (m_o + m_{\mathrm{elm}})\,\frac{v^2}{2}
\label{eq:TotalKineticEnergy}
\end{equation}
and so the observed mass of the particle would be
\begin{equation}
m \equiv m_{\mathrm{exp}} = m_o + m_{\mathrm{elm}}.
\label{eq:ObservedMass}
\end{equation}
This result (\ref{eq:ObservedMass}) was very significant, for it
appeared to allow a purely electromagnetic description of the
electron. If we set $m_o = 0$ in (\ref{eq:ObservedMass}) then from
(\ref{eq:ElectromagneticMass}) we can derive the radius of such an
electron:
\begin{equation}
R = f\,\frac{e^2}{m c^2}.
\label{eq:ElectronRadius}
\end{equation}

Much work was done on the theory of a spherical electron, Lorentz being the
most notable contributor. Lorentz was able to derive an equation of motion
for the electron, which included terms dependent upon its structure. However
the problems of how an extended charge distribution is held together (with no
cohesive forces it would explode under its own Coulomb repulsion), and the
advent of special relativity (a sphere is not Lorentz invariant) eventually caused
activity in this area to grind to a halt around 1910. The possibility of a point-like
electron was not given serious consideration either because of its divergent
self-energy.

The classical theory of the electron was to remain in this incomplete
state until Dirac made important progress in 1938 \cite{Dirac1938}. He
succeeded in deriving in a completely covariant way the equation of
motion for a point charge. His equation became known as the
\emph{Lorentz–Dirac equation}, so named because it was essentially the
equation of Lorentz with structure-dependent terms omitted. To be
sure, in Dirac's derivation the divergent self-energy of the electron
had to be absorbed into $m_o$, thus ignoring its divergent nature
and producing the first example of mass-renormalization in classical
physics. Another difficulty with Dirac's procedure was the
introduction of advanced fields in order to define a finite radiation
field close to the electron.  These advanced fields can have no direct
physical meaning for although they obey Maxwell's equations, they
propagate backwards in time, thus violating causality.

Subsequent developments in the classical theory of point charges were
characterised by the desire to rid the theory of both the
aforementioned problems.  Rohrlich \cite{Rohrlich1965} was able to
discard renormalisation, but at the expense of having to retain the
advanced fields. In 1970 Teitelboim \cite{Teitelboim1970} succeeded in
solving the converse problem: He showed that the Lorentz–Dirac
equation could be derived from retarded fields only, but to do so
still required mass renormalization.

Further developments through the 1970's centred on derivations of the
equation of motion through considerations of self-interaction. Barut in 1974
discovered a very quick way for obtaining the self-field of a point charge.
Using his method, Barut was able to give derivations of the equation of
motion in both Minkowski space and in arbitrarily curved space-times.

In this report we discuss in detail the derivations of Dirac,
Teitelboim and Barut, with a view to clearly locating the origin of
classical mass-renormalization.  The relevance of this origin to
modern-day physics cannot be over-stressed, for in all the theories of
fundamental interactions mass-renormalisation procedures are
extensively applied.

In Chapter~\ref{chap:2} we introduce relativistic electromagnetism
with a derivation of the retarded field due to a point
charge. Chapter~\ref{chap:3} is an overview of energy-momentum tensors
and their role in special relativity. Teitelboim and Dirac's
derivations of the Lorentz–Dirac equation are discussed in great
detail in Chapter~\ref{chap:4}. In Chapter~\ref{chap:5} we show how
the self-interaction of a point-charge leads to the correct equation
of motion. Our final chapter, Chapter~\ref{chap:6}, contains a
discussion of the various problems inherent in mass renormalization,
and how we may avoid them by a suitable redefinition of the field on
the world line of a point charge.

\chapter{MAXWELL'S EQUATIONS AND POINT PARTICLES}
\label{chap:2}

\section{Background}
\label{sec:2.1}

Although classical electromagnetic theory was developed to explain the
behaviour of continuous charge distributions, in reality all matter
consists of point particles (quarks, electrons, muons etc) and it is
these that carry the electromagnetic charge.  Thus the simplest
question one can ask of any theory of electromagnetism is what
predictions it makes for the behaviour of point charges. This chapter
addresses this question; beginning with an introduction to Maxwell's
equations, followed by a derivation of the field due to a charged
point particle.

\section{Maxwell's Equations}
\label{sec:2.2}

Maxwell's equations are most simply stated in the following form:
\begin{equation}
F_{\mu\nu,\rho} + F_{\nu\rho,\mu} + F_{\rho\mu,\nu} = 0
\label{eq:2.1}
\end{equation}
\begin{equation}
F^{\mu\nu}{}_{,\nu} = 4\pi j^\mu
\label{eq:2.2}
\end{equation}
where $F^{\mu\nu}$ is the Maxwell electromagnetic field tensor, and $j^\mu$ is the
current four-vector. $F^{\mu\nu}$ is antisymmetric, i.e.
\begin{equation}
F^{\mu\nu} = -F^{\nu\mu}.
\label{eq:2.3}
\end{equation}

Equation \eqref{eq:2.3} shows that $F^{\mu\nu}$ has six independent components, which
can then be interpreted as the electric and magnetic fields. In fact, for our purposes
we take $F^{\mu\nu}$ to have the following form:
\begin{equation}
F^{\mu\nu} =
\begin{bmatrix}
0 & E_1 & E_2 & E_3 \\
-E_1 & 0 & -B_3 & B_2 \\
-E_2 & B_3 & 0 & -B_1 \\
-E_3 & -B_2 & B_1 & 0
\end{bmatrix}
\label{eq:2.4}
\end{equation}
where $\mathbf{E} = (E_1, E_2, E_3)$ and $\mathbf{B} = (B_1, B_2, B_3)$ are the electric
and magnetic fields respectively.

Equation \eqref{eq:2.3} allows us to express $F_{\mu\nu}$ in terms of potentials.
Specifically, for any (suitably smooth) antisymmetric tensor $F_{\mu\nu}$ we can find
a four-vector\footnote{The reader should note that the terms ``vector'' and ``vector-field''
(and ``tensor'' and ``tensor-field'') are used interchangeably. This should not cause
confusion as it will be clear from the context which meaning is intended.} $A_\mu$, such that
\begin{equation}
F_{\mu\nu} = A_{\nu,\mu} - A_{\mu,\nu}.
\label{eq:2.5}
\end{equation}

$A_\mu$ is called the \emph{potential} of the field $F_{\mu\nu}$. Notice that \eqref{eq:2.5}
does not determine $A_\mu$ uniquely, for a transformation of the potential
\begin{equation}
A'_\mu = A_\mu + \Lambda_{,\mu}
\label{eq:2.6}
\end{equation}
where $\Lambda$ is an arbitrary scalar, does not affect the field, viz:
\begin{align*}
F'_{\mu\nu}
  &= A'_{\nu,\mu} - A'_{\mu,\nu} \nonumber \\
  &= A_{\nu,\mu} - A_{\mu,\nu}
     - \Lambda_{,\nu\mu} + \Lambda_{,\mu\nu} \nonumber \\
  &= F_{\mu\nu}.
\end{align*}

Therefore, we see that the potential $A_\mu$ does not have any direct physical significance,
it is only those quantities such as $F_{\mu\nu}$ that are invariant under the gauge
transformations \eqref{eq:2.6} that are real physical quantities. Such quantities are called
\emph{gauge invariant}.

We are able to make use of the freedom offered by gauge transformations to considerably
simplify the field equations \eqref{eq:2.2}. For example, if we add to $A_\mu$ a term
$\Lambda_{,\mu}$ such that $\Lambda^{,\mu}{}_{,\mu} = -A^\mu{}_{,\mu}$, then our new potential
satisfies
\begin{equation}
A^\mu{}_{,\mu} = 0
\label{eq:2.7} \quad\textit{(Lorenz gauge condition)}
\end{equation}
With this condition on our potential, \eqref{eq:2.2} becomes
\begin{equation*}
A^{\mu,\nu}{}_{,\nu} = -4\pi j^\mu
\end{equation*}
or
\begin{equation}
\square A^\mu = -4\pi j^\mu
\label{eq:2.8}
\end{equation}
where $\square \equiv \partial_\nu \partial^\nu$ is the d'Alembertian operator. Even with
the choice of Lorenz gauge \eqref{eq:2.7}, $A^\mu$ is not determined uniquely and in fact
still admits a gauge transformation of the form $A'^\mu = A^\mu + f^{,\mu}$ where $f$ satisfies
the d'Alembertian equation $\square f = 0$.

\section{Lienard-Wiechert potentials}
\label{sec:2.3}

\begin{figure}[!t]
  \vspace*{-1.8cm}
  \centering
  \includegraphics[width=0.65\textwidth]{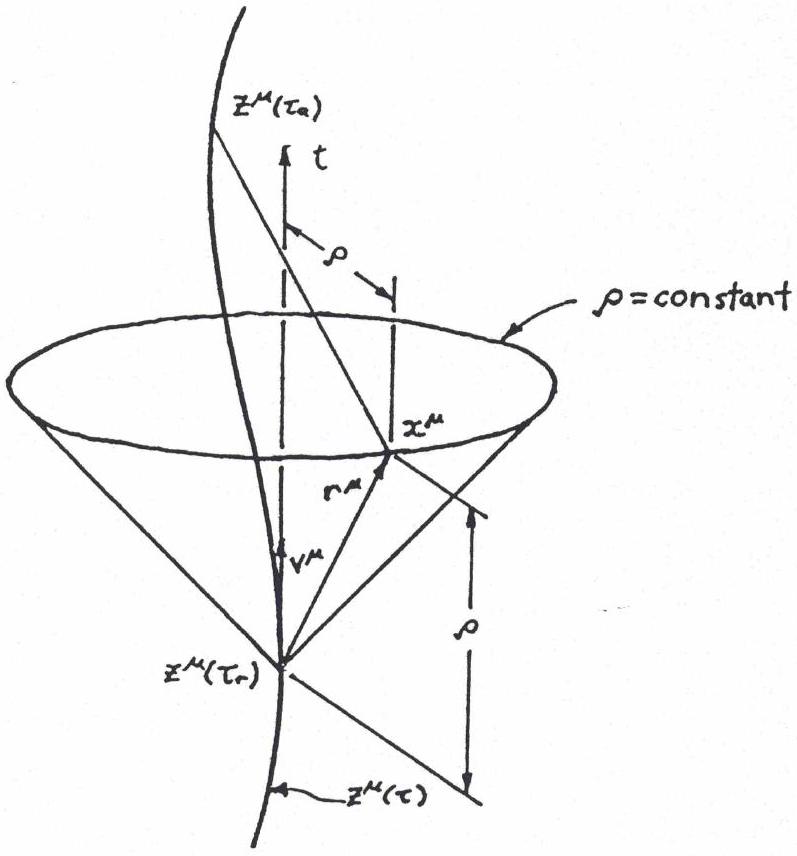}
  \caption{Light-cone and World-line of a point charge: $\tau_a$ and $\tau_r$
  are the advanced and retarded proper-times corresponding to $x^\mu$; $v^\mu$
  is the charge's four-velocity; $\rho$ is the spatial distance between the
  retarded point of the charge and the point $x^\mu$, in the instantaneous
  retarded rest-frame (i.r.r.f).}
  \label{fig:2.1}
\end{figure}

We are now interested in obtaining the field at an arbitrary point of space
time due to the influence of a point charge. To do this we must solve
\eqref{eq:2.8} for $A^\mu$ in the particular instance of $j^\mu$ being the
four-current of a point charge. With reference to Figure~\ref{fig:2.1}, we
denote the world line of the charge by $z^\mu(\tau)$ where $\tau$ is the
proper time experienced by the charge. The four-velocity of the charge,
$v^\mu(\tau)\equiv \dot{z}^\mu(\tau)$, satisfies
\begin{equation}
  v^2 = -1
  \label{eq:2.9}
\end{equation}
by virtue of the definition of $\tau$.

To determine the form of $j^\mu$ we observe that in any reference frame with
coordinates $(t,\mathbf{x})$ the charge distribution for a charge of magnitude
$e$ will be given by
\begin{equation*}
  \rho(t,\mathbf{x}) = e \delta^3(\mathbf{x} - \mathbf{z}(t))
\end{equation*}
and hence in such a frame the current distribution is
\begin{equation*}
  j^\mu(t,\mathbf{x}) = e \delta^3(\mathbf{x} - \mathbf{z}(t))
  \frac{\partial z^\mu}{\partial t}
\end{equation*}
which, in more covariant language, can be expressed as\footnote{Note that this
form for $j^\mu$ yields immediately the required conservation law:
\[j^\mu{}_{,\mu} = 0\].}
\begin{equation}
  j^\mu = e \int \delta^4(x - z(\tau)) v^\mu d\tau.
  \label{eq:2.10}
\end{equation}

This form of $j^\mu$ may now be substituted into \eqref{eq:2.8} and a solution
found for the four-vector potential $A^\mu$ by the method of Green's
functions. The interested reader is referred to \cite{Barut1964}
for a full description of this procedure. We simply quote the final result,
which is:
\begin{equation}
  A^\mu(x) = e \left[ \frac{v^\mu}{\rho} \right]_{\mathrm{ret}}
  \label{eq:2.11}
\end{equation}
where for notational convenience we have defined the scalar quantity $\rho$ by
\begin{equation}
  \rho = - (x^\alpha - z^\alpha(\tau)) v_\alpha(\tau).
  \label{eq:2.12}
\end{equation}
The subscript $\mathrm{ret}$ indicates that quantities dependent on $\tau$
(e.g $v^\mu$, $z^\mu$ etc) are to be evaluated at the proper time $\tau_r$,
where $\tau_r$ is the \emph{retarded} solution of the light-cone
equation\footnote{The existence and uniqueness of retarded world-line point
$z(\tau_r)$ is guaranteed by the fact that $v^\mu(\tau)$ is everywhere
timelike, as long as we exclude the pathological instance of a world-line that
asymptotically approaches the light-cone as $\tau\rightarrow -\infty$.}
\begin{equation}
  (x - z(\tau_r))^2 = 0.
  \label{eq:2.13}
\end{equation}

Equation \eqref{eq:2.13} will in general have two solutions for any
given space-time point $x$; the \emph{retarded} solution lies on the
past light-cone from $x$, and the \emph{advanced} solution lies on the
future light-cone (see Figure~\ref{fig:2.1}). We take the retarded
solution as the physical one, reflecting the empirical fact that
electromagnetic influences propagate forwards, not backwards, in
time. The dependence of $A^\mu$ upon the retarded world-line point
$z(\tau_r)$ also makes explicit the fact that electromagnetic
influences propagate at the speed of light. In future we will make use
of the following abbreviation:
\begin{equation}
  r^\alpha = x^\alpha - z^\alpha(\tau_r),
  \label{eq:2.14}
\end{equation}
i.e $r^\alpha(x)$ is the null vector connecting space-time point $x$ with its
retarded world-line point, $z(\tau_r)$.

To gain some insight into the physical significance of
$\rho = \rho(x,\tau_r)$ we can look at this quantity in the rest frame of the
particle at proper time $\tau_r$. In this frame the four velocity, $v^\alpha$,
has components $(1,0,0,0)$ and since $r^\alpha$ is a null vector, we find from
\eqref{eq:2.12}
\begin{equation}
  \rho(x,\tau_r) = \left| \mathbf{x} - \mathbf{z}(\tau_r) \right|.
  \label{eq:2.15}
\end{equation}
Thus we identify $\rho$ with the spatial distance between the field point $x$
and the retarded position of the charge, as viewed from the rest frame of the
particle at proper time $\tau_r$. This point is illustrated in
Figure~\ref{fig:2.1}.

\section{Differentiation of retarded functions}
\label{sec:2.4}

In order to evaluate the field tensor $F^{\mu\nu}$ we must
differentiate the potential~\eqref{eq:2.11} according
to~\eqref{eq:2.5}. This task is non-trivial as $A^\mu$ is dependent
not only on the field point $x$, but also on the retarded quantities
$v(\tau_r)$ and $z(\tau_r)$. Thus before embarking on a calculation of
$F^{\mu\nu}$ we derive the techniques necessary for differentiation of
functions dependent on retarded quantities.

If we make a variation in the field point, $x^\mu \rightarrow x^\mu +
\delta x^\mu$, then the retarded proper time corresponding to $x$ will
also undergo a variation $\tau_r \rightarrow \tau_r + \delta
\tau_r$. This variation must occur in such a manner as to leave $r^\mu
= x^\mu - z^\mu(\tau_r)$ still a null vector, i.e
\begin{align}
\delta\left[(x - z(\tau_r))^2\right] &= 0 \notag\\
\Rightarrow (\delta x^\mu - \delta z^\mu(\tau_r))(x_\mu - z_\mu(\tau_r)) &= 0
\label{eq:2.16}
\end{align}
Writing
\begin{equation*}
\delta z^\mu(\tau_r) = v^\mu(\tau_r)\,\delta\tau_r
\end{equation*}
and substituting this expression into~\eqref{eq:2.16} we find
\begin{equation*}
\rho\,\delta\tau_r + r^\mu \delta x_\mu = 0
\end{equation*}
or equivalently
\begin{equation}
  \frac{\partial \tau_r}{\partial x_\mu} = -\,\frac{r^\mu}{\rho}.
  \label{eq:2.17}
\end{equation}
Thus if $g(\tau_r)$ is any function dependent only on the retarded time, $\tau_r$,
then
\vspace*{-2mm}
\begin{equation}
  \partial_\mu g(\tau_r) = -\frac{r_\mu}{\rho}\left.\frac{d}{d\tau} g(\tau)\right|_{\tau=\tau_r}.
  \label{eq:2.18}
\end{equation}
\vspace*{-5mm}Using~\eqref{eq:2.18} we find the following identities:
\begin{align}
  \partial_\mu r^\mu  &= 3 \notag\\
  \partial_\mu r^\nu  &= \delta^\nu_\mu + \frac{r_\mu}{\rho} v^\nu \notag\\
  \partial_\mu v^\mu  &= -a_r \notag\\
  \partial_\mu v^\nu  &= -\frac{r_\mu}{\rho} a^\nu \label{eq:2.19} \\
  \partial_\mu a^\mu  &= -\dot{a}_r \notag\\
  \partial_\mu a^\nu  &= -\frac{r_\mu}{\rho} \dot{a}^\nu \notag \\
  \partial_\mu \rho   &= a_r\,r_\mu - v_\mu + \frac{r_\mu}{\rho} \notag
\end{align}
where $\cdot{} \equiv \frac{d}{d\tau}$, $a^\mu = \dot{v}^\mu$ and $a_r$, $\dot{a}_r$
stand for $a^\mu r_\mu / \rho$ and $\dot{a}^\mu r_\mu / \rho$ respectively.

We also have the following identities derived from~\eqref{eq:2.9}
\begin{gather}
v^\mu a_\mu = 0 \label{eq:2.20} \\
v^\mu \dot{a}_\mu = -a^2. \label{eq:2.21}
\end{gather}

Substituting~\eqref{eq:2.11} into~\eqref{eq:2.5} and using the above
identities we obtain the formula for the retarded electromagnetic
field tensor due to the presence of a point charge:
\begin{equation}
F^{\mu\nu} = \frac{e}{\rho^3}v^{[\mu} r^{\nu]}
+ \frac{e}{\rho^2}
\bigl(a_r\,v^{[\mu} r^{\nu]} + a^{[\mu} r^{\nu]}\bigr)
\label{eq:2.22}
\end{equation}
with
\begin{equation}
a^{[\mu} b^{\nu]} = a^\mu b^\nu - a^\nu b^\mu.
\label{eq:2.23}
\end{equation}

\chapter{MOMENTUM AND RADIATION}
\label{chap:3}

\section{Background}
\label{sec:3.1}

Classically, the physical behaviour of a system of interacting mass points is
determined by the law of conservation of four-momentum: the system must
evolve in such a way that its total four-momentum is unchanging. We know,
however, that particles carrying electric charge experience a force when in
the presence of an electromagnetic field. Thus electromagnetic fields are
capable of imparting four-momentum to charged particles. Hence in order
for the law of conservation of four-momentum to remain valid for systems of
interacting particles and fields, we see that the electromagnetic field must
carry some four-momentum of its own, and when interacting with charged
particles some of this four-momentum is transferred to the particle.

The usual way of treating conservation-of-momentum laws in special
relativity is through the use of an appropriate \textit{energy-momentum tensor}. In
this chapter we will discuss the origin and importance of such tensors and
give a few simple examples. The concept of electromagnetic radiation will
be introduced, and it will be shown that there is part of the field of a point
charge that can thought of as radiating away from the charge.

\section{Energy-momentum tensor}
\label{sec:3.2}

To begin with we motivate the use of energy-momentum tensors as a means
of dealing with systems of interacting particles and fields. This is best
achieved by first considering the concept of a four-current in special relativity.

The definition of four-current is very straightforward: any vector field,
$J^\mu$, satisfying $J^\mu{}_{,\mu}=0$ is called a four-current. With any four-current is
associated a conserved charge, which we will denote by $Q$. $Q$ is a scalar and
is given by the equation
\begin{equation}
Q=\int_S j^\mu dS_\mu .
\label{eq:3.1}
\end{equation}
where $S$ is any infinite space-like surface (i.e.\ the normal to $S$ is everywhere
timelike). To demonstrate that $Q$ is a conserved quantity we must show that
its definition~\eqref{eq:3.1} is independent of the surface $S$. To do this we consider
$Q$ defined by two separate surfaces $S$ and $S'$, and calculate $Q\big|_S - Q\big|_{S'}$.
If we let $V$ be the space-time volume bounded by $S$ and $S'$, and we denote the
boundary of $V$ by $\partial V$, then
\vspace*{-2mm}
\begin{align*}
  Q\big|_S - Q\big|_{S'}
  &= \int_S j^\mu dS_\mu - \int_{S'} j^\mu dS'_\mu \\
  &= \int_{\partial V} j^\mu dS_\mu \\
  &= \int_V j^\mu_{,\mu} d^4v \quad\text{(by Gauss' theorem)} 
\end{align*}
\[=0\]
Thus $Q$ is independent of $S$ and so is conserved\footnote{In
particular $Q$ is invariant under time translations of the surface $S$
and hence is a constant of the motion.}.

By analogy with the considerations above, if we have conservation of
total four-momentum $P^\alpha$ then there should exist an associated
four-current $T^{\alpha\beta}$ for each $\alpha=0,1,2,3$, satisfying
$T^{\alpha\beta}{}_{,\beta}=0$, and such that
\begin{equation}
P^\alpha = \int_S T^{\alpha\beta} dS_\beta .
\label{eq:3.2}
\end{equation}
Again $P^\alpha$ so defined will be independent of the choice of
$S$. To ensure conservation of four-momentum in all frames we also
require $T^{\alpha\beta}$ to be a tensor. Furthermore, conservation of
angular momentum $M^{\alpha\beta}=x^\alpha P^\beta - x^\beta P^\alpha$
forces $T^{\alpha\beta}$ to be symmetric. Any tensor satisfying the
above requirements is called an \emph{energy-momentum tensor}.

Having justified the existence of $T^{\alpha\beta}$ we would like to have some idea
of the physical significance of its components. To this end we examine the
individual components of $P^\alpha$.

$\alpha=0$. The zeroth component of four-momentum $P^0$ is usually
taken to be the total energy of the system, $E$. Thus, from
~\eqref{eq:3.2}, $T^{00}$ is clearly the energy density, and $T^{0i}$
is the energy flux density.

$\alpha=i$. $P^i$ is the $i^{\text{th}}$ component of the total
momentum and so referring to~\eqref{eq:3.2} again we see that $T^{i0}$ is the
density of the $i^{\text{th}}$ component of momentum.  $T^{ij}$ then
represents the flux density in the $j^{\text{th}}$ direction of the
$i^{\text{th}}$ component of momentum. Thus $T^{ii}$, being
$i$-momentum flux in the $i$-direction, is interpreted as the pressure
exerted by the field, while components with $i\neq j$ are shearing
stresses.

It is possible to define energy-momentum tensors for both fields and
general material distributions. To give some idea of how this may be
done we present a few examples.

\vspace{\baselineskip}

\textbf{1) Energy-momentum tensor for a pressure-free, inviscid fluid.}

\vspace{\baselineskip}

The fluid is described by a four-velocity $V^\mu(x)$ and a local rest mass
density $\rho(x)$ at each point $x$. To find $T^{\mu\nu}$ in an arbitrary frame we first look
at $T^{\mu\nu}$ in the local instantaneous rest frame (i.r.f) at each point. In this
frame $V^\mu=(1,0,0,0)$ and $E=$ local rest-energy density $=\rho$. As we are in
the i.r.f, there will be no momentum density and so $T^{0i}=T^{i0}=0$. Also the
restriction to a pressure-free, inviscid fluid means that $T^{ij}=0$. Thus the
only non-zero component is $T^{00}=\rho$. Noting the form of $V^\mu$ we can write
this as:
\begin{equation}
T^{\mu\nu}=\rho V^\mu V^\nu .
\label{eq:3.3}
\end{equation}

As this is a tensor equation it must be true in all frames. (Note the
explicit symmetry in $T^{\mu\nu}$.)

\vspace{\baselineskip}

\textbf{2) Energy-momentum tensor for a point particle.}

\vspace{\baselineskip}

Here we assume that the particle is described by a world-line $z^\mu(\tau)$ with
four-velocity $v^\mu(\tau)$ as in the previous chapter. We also assume that the
particle is stable and hence does not have any internal stress. Working
again in the instantaneous rest frame of the particle, we write $m_o$ for the
rest-mass. Then the total rest-energy is $m_o$ and since this is concentrated at
the point $z(\tau)$, the rest-energy density in the i.r.f is $E=m_o \delta^3(\mbf{x}-\mbf{z}(t))$.
In a more covariant form (\textit{cf.}~\eqref{eq:2.10}) this becomes
\begin{equation*}
E=\int m_o \delta^4(x-z(\tau))\, d\tau .
\end{equation*}
As in the previous example this is the only contributing component of $T^{\mu\nu}$
and so by analogy we have immediately:
\begin{equation}
T^{\mu\nu}=\int m_o v^\mu v^\nu \delta^4(x-z(\tau))\, d\tau
\label{eq:3.4}
\end{equation}

$T^{\mu\nu}{}_{,\nu}=0$ implies, after some algebra, that $m_o a^\mu=0$, i.e.\ conservation of
four-momentum $\equiv v^\mu=$ constant in this case.

It is instructive to calculate the four-momentum associated with this
energy-momentum tensor. Substituting~\eqref{eq:3.4} into~\eqref{eq:3.2} we have:
\begin{equation*}
P^\mu=\int_S \left( \int m_o v^\mu v^\nu \delta^4(x-z(\tau))\, d\tau \right) dS_\nu
\end{equation*}
Now if we take our space-like surface $S$ to be the rest three-space of the
particle at each proper-time, $\tau$, then $v^\nu$ and $dS^\nu$ will be parallel four-vectors,
i.e.\ $v^\nu$ will be normal to the surface $S$. Thus our integral will reduce to:
\begin{equation*}
P^\mu=\int_V m_o v^\mu \delta^4(x-z(\tau))\, d^4x
\end{equation*}
or
\begin{equation*}
P^\mu = m_o v^\mu
\end{equation*}
as we would hope.

The above two examples serve to introduce energy-momentum tensors for
matter distributions. Our main interest, however, is in the
energy-momentum tensor of an electromagnetic field. Unfortunately it
is not possible to apply the simple techniques used above to derive an
expression for the electromagnetic energy-momentum tensor, and so more
elaborate (and more general) methods must be found. The simplest and
neatest way known to us for deriving any energy-momentum tensor is
through the use of a variational principle in general
relativity. However it is beyond the scope of this report to go into
the details of this technique here, and so we shall simply state the
result:
\begin{equation}
T^{\mu\nu}=-\frac{1}{4\pi}\left(F^{\mu\alpha}F_\alpha{}^\nu + \frac14 F^{\alpha\beta}F_{\alpha\beta} g^{\mu\nu}\right).
\label{eq:3.5}
\end{equation}
For a full derivation see~\cite{LandauLifshitz80}.

To get some intuitive idea of what $T^{\mu\nu}$ represents physically
we can evaluate its various components in terms of the electric and
magnetic fields, $\mbf{E}$ and $\mbf{B}$. Substituting our
form~\eqref{eq:2.4} for $F^{\mu\nu}$ into~~\eqref{eq:3.5} we find that
\begin{align*}
T^{00}&=\frac{\mbf{E}^2+\mbf{B}^2}{8\pi} \qquad \textit{Energy Density},\\
T^{0i}&=\frac{(\mbf{E}\times\mbf{B})^i}{4\pi} \qquad \textit{Poynting Vector},\\
T_{ij}&=\frac{1}{4\pi}(-E_i E_j - B_i B_j + \tfrac12 (\mbf{E}^2+\mbf{B}^2)\delta_{ij}) \qquad \textit{Maxwell stress tensor}.
\end{align*}
We see that by expressing $T^{\mu\nu}$ in terms of $\mbf{E}$ and $\mbf{B}$ the origin of the classical
formulae for energy density, energy flux density (Poynting vector) and the
Maxwell stress tensor become apparent.

Given that $T^{\mu\nu}$ defined by equation~\eqref{eq:3.5} above is
supposed to represent an energy-momentum tensor, we should like to
find an expression for its divergence. Applying
equations~\eqref{eq:2.2} and~\eqref{eq:2.1} we have
\begin{equation}
T^{\mu\nu}{}_{,\nu}=-F^{\mu\nu}j_\nu .
\label{eq:3.6}
\end{equation}
Thus we see that in regions of space-time free of sources ($j^\mu=0$), the
divergence of $T^{\mu\nu}$ is zero, and so four-momentum in the field is conserved.
However, if $j^\mu\neq 0$ then four-momentum in the field is not conserved, and
is in fact `pumped in' by the current $j^\mu$.

\section{Radiation}
\label{sec:3.3}

It is a well known fact that a charged particle radiates
electromagnetic energy when accelerated. The radiation is propagated
in waves, emitted from the charge and travelling with the speed of
light. This radiation behaviour of the field of a charge is in
contradistinction to the Coulomb field of the charge, which cannot be
thought of as radiating away and in a sense is ``bound'' to the
charge. To see how these separate phenomena arise we examine more
closely the electromagnetic field due to a point charge,~\eqref{eq:2.22}.

The electromagnetic field tensor decomposes naturally into two parts
\begin{equation}
F^{\mu\nu} = F^{\mu\nu}_{I} + F^{\mu\nu}_{II}
\label{eq:3.7}
\end{equation}
where
\begin{equation}
F^{\mu\nu}_{I} = \frac{e}{\rho^{3}} v^{[\mu} r^{\nu]}
\label{eq:3.8}
\end{equation}
is the near or velocity field which is responsible for the Coulomb
behaviour of the charge, and
\begin{equation}
F^{\mu\nu}_{II} = \frac{e}{\rho^{2}} (a_{r} v^{[\mu} r^{\nu]} + a^{[\mu} r^{\nu]})
\label{eq:3.9}
\end{equation}
is the far or acceleration field, responsible for radiation.

To compare the contributions to the field at space-time point $x$ of
these two components of $F^{\mu\nu}$ we view them in the instantaneous
retarded rest frame (i.r.r.f) of the charge, i.e in the rest frame of
the charge at the point of intersection between its world-line and the
past light-cone from $x$. In this frame we have the following
definitions and results (\textit{cf.}~\eqref{eq:2.14}, \eqref{eq:2.15}):
\begin{align}
r^{\mu} &= R(1,\mbf{n}) \quad (\mbf{n} \text{ is a unit vector},) \label{eq:3.10} \\
v^{\mu} &= (1,0) \label{eq:3.11} \\
a^{\mu} &= (0,\mbf{a}) \label{eq:3.12} \\
\rho    &= R     \label{eq:3.13} \\
a_r     = \frac{a^{\mu} r_{\mu}}{\rho} &= \mbf{a} \cdot \mbf{n}\label{eq:3.14}
\end{align}
We are primarily interested in an order of magnitude estimate for the
fields. In particular we would like to compare the $R$ dependence of
each field as $R$ is the spatial distance between the charge and the
field point $x$ (in the i.r.r.f). Looking first at $F^{\mu\nu}_{I}$,
the near field, and applying~\eqref{eq:3.10},~\eqref{eq:3.11}
and~\eqref{eq:3.13} we see that
\begin{equation}
|F^{\mu\nu}_{I}| \sim \frac{1}{R^{2}}
\label{eq:3.15}
\end{equation}
while for the far field we have
\begin{equation}
|F^{\mu\nu}_{II}| \sim \frac{1}{R}.
\label{eq:3.16}
\end{equation}
So $F^{\mu\nu}_{I}$ and $F^{\mu\nu}_{II}$ justify their designation as
the near and far fields, for $F^{\mu\nu}_{I}$ clearly will dominate
the total field near to the charge, and $F^{\mu\nu}_{II}$ dominates in
regions far removed from the charge. We note here also that
$F^{\mu\nu}_{II}$ is proportional only to the four-acceleration and
not the four-velocity, and so a uniformly moving charge will not have
a far field.

In order to see that there will be some of the field of the charge
that is radiated away at infinity, it is necessary to examine the
contribution each of the near and far fields make to the
energy-momentum tensor,~\eqref{eq:3.5}. If we write $T^{\mu\nu}$ in
terms of the near and far fields we find
\begin{equation*}
T^{\mu\nu} = T^{\mu\nu}_{I} + T^{\mu\nu}_{I,II} + T^{\mu\nu}_{II},
\end{equation*}
where $T^{\mu\nu}_{I}$ and $T^{\mu\nu}_{II}$ are the tensors obtained
when~\eqref{eq:3.5} is evaluated with the fields $F^{\mu\nu}_{I}$ and
$F^{\mu\nu}_{II}$ respectively, and $T^{\mu\nu}_{I,II}$ is the result
of interference between the two fields. From~\eqref{eq:3.15}
and~\eqref{eq:3.16} we can determine the $R$ dependence of the various
components of $T^{\mu\nu}$, viz:
\begin{align}
\left|T^{\mu\nu}_{I}\right|    &\sim \frac{1}{R^{4}}\notag\\
\left|T^{\mu\nu}_{I,II}\right| &\sim \frac{1}{R^{3}}\label{eq:3.17}\\
\left|T^{\mu\nu}_{II}\right|   &\sim \frac{1}{R^{2}}\notag
\end{align}
\enlargethispage{0.7\baselineskip}
To calculate the four-momentum radiation rate (still working in the
i.r.r.f) we break the problem into two parts: a calculation of the
energy radiation rate, and a calculation of the ordinary
three-momentum radiation rate. For now we will look at the former
problem. To calculate the energy radiation rate $(dE_{rad}/dt)$ we
integrate the energy flux density (Poynting vector) over the surface
of a sphere centered on the charge, and let the radius of the sphere
tend to infinity. The components of the energy flux density are
$T^{0i}$ and since the surface area of the sphere is $4\pi R^{2}$,
where $R$ is the radius of the sphere, we have
\begin{equation}
  \frac{dE_{rad}}{dt} \sim |T^{0i}| R^{2}.
  \label{eq:3.18}
\end{equation}

By substituting the expressions for the $R$ dependence of the various
components of $T^{\mu\nu}$~\eqref{eq:3.17} into~\eqref{eq:3.18} we
find that the only element of $T^{\mu\nu}$ that can contribute to
radiation at infinity is $T^{\mu\nu}_{II}$.

It must be stressed at this point that the preceding arguments are by
no means a rigorous justification for the existence of electromagnetic
radiation for an accelerated point charge, but are intended merely as
a guide to further our intuitive understanding. In particular, to
write an equation like~\eqref{eq:3.18} we are implicitly assuming that
the angular dependence of $T^{\mu\nu}$ in the i.r.r.f is not such that
when integrated over the surface of the sphere it yields the value
zero. We have, however, demonstrated that $T^{\mu\nu}_{I}$ and
$T^{\mu\nu}_{I,II}$ certainly cannot contribute to radiation, and so
any radiative effects in our theory must come solely from the far
field, $F^{\mu\nu}_{II}$. To demonstrate fully the phenomena of
radiation we have to show that at large distances (large $R$) the
electromagnetic field approximates a plane wave, and in addition we
need to calculate explicitly the radiated four-momentum. This we do in
the following section.

\subsection{Plane Waves}
\label{sec:3.3.1}
{
\setlength{\abovedisplayskip}{4pt plus 2pt minus 2pt}%
\setlength{\belowdisplayskip}{4pt plus 2pt minus 2pt}%
To find the behaviour of the electromagnetic energy-momentum tensor at
large distances from the charge we need to evaluate
$T^{\mu\nu}_{II}$. This is achieved by inserting~\eqref{eq:3.9} into~\eqref{eq:3.5}:
\begin{equation}
  T^{\mu\nu}_{II} = -\frac{e^{2} r^{\mu} r^{\nu}}{4\pi \rho^{4}} \big( (a_{r}^{2} - a^{2}) \big).
  \label{eq:3.19}
\end{equation}
We now evaluate~\eqref{eq:3.19} in the i.r.r.f of the charge,
using~\eqref{eq:3.10}--\eqref{eq:3.14}:
\begin{align}
  T^{00}_{II} &= -\frac{e^{2}}{4\pi R^{2}} \left((\mbf{a}\cdot \mbf{n})^{2} - \mbf{a}^{2}\right)
  \label{eq:3.20} \\
  T^{0i}_{II} &= -\frac{e^{2} n^{i}}{4\pi R^{2}} \left((\mbf{a}\cdot \mbf{n})^{2} - \mbf{a}^{2}\right)
  \label{eq:3.21} \\
  T^{ij}_{II} &= -\frac{e^{2} n^{i} n^{j}}{4\pi R^{2}} \left((\mbf{a}\cdot \mbf{n})^{2} - \mbf{a}^{2}\right).
  \label{eq:3.22} 
\end{align}

To begin with we evaluate the energy radiation rate at infinity. As
mentioned above this involves integrating the Poynting vector over the
surface of a sphere centred on the retarded point of the charge. We
will denote by $S_{II}$ the Poynting vector due to $T^{\mu\nu}_{II}$, and
so we have
\begin{equation}
  S^{i}_{II} = T^{0i}_{II} = -\frac{e^{2} n^{i}}{4\pi R^{2}} \left((\mbf{a}\cdot\mbf{n})^{2} - \mbf{a}^{2}\right).
  \label{eq:3.23}
\end{equation}
Thus the energy is being carried away in direction $\mbf{n}$ as is
required for radiation. In polar coordinates we have
\begin{equation*}
  \mbf{n} = (\sin\theta \cos\phi, \sin\theta \sin\phi, \cos\theta).
\end{equation*}
Substituting this into~\eqref{eq:3.23} and integrating over a sphere
of radius $R$, with surface element $d\mbf{A} = \mbf{n}R^{2}\sin\theta
d\theta d\phi$ centred on the charge, yields the energy radiation
rate.
\begin{align}
  \frac{dE_{rad}}{dt}
  &= \lim_{R\to\infty}\int_{\text{Sphere}}\mbf{S}_{II} \cdot d\mbf{A} \notag\\
  &= -\frac{e^{2}}{4\pi}\int_{0}^{2\pi}d\phi\int_{0}^{\pi}\sin\theta d\theta\left((\mbf{a}\cdot \mbf{n})^{2} - \mbf{a}^{2} \right) \notag\\
  &=\frac{2}{3} e^{2} \mbf{a}^{2},\label{eq:3.24}
\end{align}
}
where we have made use of the identity
{
\setlength{\abovedisplayskip}{4pt plus 2pt minus 2pt}%
\setlength{\belowdisplayskip}{4pt plus 2pt minus 2pt}%
\begin{equation}
\int_{0}^{2\pi} d\phi \int_{0}^{\pi} \sin\theta d\theta\, n_{i} n_{j} = \frac{4\pi}{3} \delta_{ij}.
\label{eq:3.25}
\end{equation}
It is interesting to observe that the limit as $R$ tends to infinity
never had to be taken in the derivation~\eqref{eq:3.24} because the
energy radiation rate turns out to be independent of the radius
$R$. This property will be investigated further in the following
chapter.

To complete our calculation of the radiation-rate of four-momentum we
must evaluate the rate at which three-momentum is radiated away by the
charge. This is achieved by using the same technique as was employed
to calculate the energy radiation rate, namely: integrate the momentum
flux density $T^{ij}$ over the surface of a sphere of radius $R$
centered on the charge, and take the limit as $R$ tends to infinity.
\begin{align}
  \frac{dP^{i}_{rad}}{dt}
  &=\lim_{R\to\infty}\int_{\text{Sphere}}T^{ij} n_{j} dA\notag\\
  &=-\frac{e^{2}}{4\pi}\int_{0}^{2\pi} d\phi\int_{0}^{\pi}\sin\theta d\theta\, n_{i} \left( (\mbf{a}\cdot \mbf{n})^{2} - \mbf{a}^{2} \right)\notag\\
  &=0,\label{eq:3.26}
\end{align}
where we have used the identities
\begin{equation*}
\int_{0}^{2\pi} d\phi \int_{0}^{\pi} \sin\theta d\theta\, n_{i}
=
\int_{0}^{2\pi} d\phi \int_{0}^{\pi} \sin\theta d\theta\, n_{i} n_{j} n_{k}
=
0.
\end{equation*}
Thus there is no three momentum radiated away in the instantaneous
rest frame (i.r.f) of the charge. If we now put the energy radiation
rate and the three-momentum radiation rate together in the radiated
four-momentum we have (for the i.r.f):
\begin{equation*}
  \frac{dP^{\mu}_{rad}}{dt}=\left( \frac{2}{3} e^{2} \mbf{a}^{2}, 0, 0, 0 \right),
\end{equation*}
}
which, by virtue of the fact that $v^{\mu} = (1,0,0,0)$, $a^{\mu} =
(0,\mbf{a})$ and $dP^{\mu}/dt = dP^{\mu}/d\tau$ in the i.r.f, can be
written as
\begin{equation}
  \frac{dP^{\mu}_{rad}}{d\tau}=\frac{2}{3} e^{2} a^{2} v^{\mu}.
  \label{eq:3.27}
\end{equation}
As this is a tensor equation, it must be the four-momentum radiation
rate in an arbitrary frame.

To complete our justification that the electromagnetic field at
infinity is a plane wave we evaluate $\mbf{E}$ and $\mbf{B}$ in the i.r.r.f. We
find
\begin{itemize}
\item (i) $\mbf{E} \perp \mbf{B}$
\item (ii) $\mbf{E} \times \mbf{B} \propto \mbf{n}$
\end{itemize}
indicating that as $R \rightarrow \infty$, the field approaches that of a plane wave.

\section{Equation of Motion}\label{sec:3.4}

Having found that an accelerating charged particle radiates
electromagnetic four-momentum, according to~\eqref{eq:3.27}, we are
prompted to inquire what consequences this may have for the equation
of motion of a point charge. Clearly the usual form of the equation of
motion, the \textit{Lorentz Force Equation}
\begin{equation}
  ma^{\mu} = eF^{\mu\nu}_{ext} v_{\nu}
  \label{eq:3.28}
\end{equation}
which governs the behaviour of a point charge moving in an external
field $F^{\mu\nu}_{ext}$, cannot be valid because it predicts constant
elliptical orbits for the motion of the charge in a uniform magnetic
field, which would imply that the charge can radiate continuously
without losing any energy.

As a first naive attempt to rectify this problem we might try to take
account of the radiated four-momentum by demanding that the charge
loses four-momentum at the rate at which it is radiating away. Thus we
might expect the following equation to be true:
\begin{equation}
  ma^{\mu} = eF^{\mu\nu}_{ext} v_{\nu} - \frac{2}{3} e^{2} a^{2} v^{\mu}.
  \label{eq:3.29}
\end{equation}
However this cannot be correct, because if we contract both sides with
$v_{\mu}$ we are left with
\begin{equation*}
\mbf{a}^{2} = 0.
\end{equation*}

Thus a more sophisticated approach to the problem is required and we
pursue this in the next chapter, where various techniques for
obtaining the equation of motion of a radiating charge are examined.

\chapter{EQUATIONS OF MOTION}
\label{chap:4}

\section{Background}
\label{sec:4.1}

In this chapter we concern ourselves with the problem of finding the
equation of motion obeyed by a point charged particle. We examine the
earliest attempt to derive this equation from considerations of energy
balance in the field of the charge, given by Dirac \cite{Dirac1938} in
1938. Dirac's method contains many undesirable features, perhaps the
worst being the use of advanced solutions to Maxwell's equations
(\textit{cf.}\ Chapter~\ref{chap:2}). However it is both physically
and historically instructive to discuss Dirac's paper, the former
because it is in this paper that classical \emph{mass-renormalization}
makes its first appearance as a means of dealing with the infinite
self-energy of a point-charge, and the latter because it was not until
1970 that the problems associated with advanced fields were removed
from Dirac's theory. In the intervening years very little effort was
put into this most fundamental problem, which is surprising given the
obvious limitations of existing theory at the time. Perhaps with the
apparent success of Quantum Electrodynamics there seemed little point
in re-examining classical electrodynamics, particularly when it failed
to explain even the simplest of atomic phenomena.

In 1970 Teitelboim \cite{Teitelboim1970} showed how to calculate the
equation of motion without resorting to the use of advanced fields. In
the second part of the chapter we give a detailed discussion of this
paper, which carries the separation of the electromagnetic field
into near and far parts, as in the preceding chapter, to its logical
end.

\section{The Dirac Procedure}
\label{sec:4.2}

We follow Dirac's \cite{Dirac1938} derivation of the equation of motion of a radiating
point charge. This procedure is characterised by Dirac's desire to place the
advanced and retarded solutions to Maxwell's equations on an equal footing
in Electrodynamics.

\subsection{The Field}
\label{sec:4.2.1}

The world line of our charge is given by $z_\mu(\tau)$ as in
\S\ref{sec:2.3}. We denote the field of~\eqref{eq:2.22} by
$F^{\mu\nu}_{\text{ret}}$. This is the usual retarded solution to
Maxwell's equations, so named because the field at any space-time
point is a function of the retarded position of the charge. In
addition to this solution there is also the advanced solution,
$F^{\mu\nu}_{\text{adv}}$, which gives the field as a function of the
advanced position of the charge (see Figure~\ref{fig:2.1}).

We wish to consider the behaviour of the charge under the action of
external fields, and so we assume that there are electromagnetic waves incident
upon the charge, whose fields we shall denote by $F^{\mu\nu}_{\text{in}}$. Thus, calling the actual
field $F^{\mu\nu}_{\text{act}}$, we have
\begin{equation}
F^{\mu\nu}_{\text{act}} = F^{\mu\nu}_{\text{ret}} + F^{\mu\nu}_{\text{in}}
\label{eq:4.1}
\end{equation}

Since the advanced and retarded fields are derived from the same
equations we would expect them to play symmetrical roles in any theory
of charged particles\footnote{ This is Dirac's justification for
introducing the advanced fields. However we cannot entirely agree with
the idea that advanced and retarded fields should play symmetrical
roles, for it is clear from everyday phenomena that the advanced
solutions play no part at all in electrodynamics. If they did,
causality would be violated on a macroscopic scale.  Thus, at some
stage in the derivation of the equation of motion we must introduce
this asymmetry, and in order not to violate causality this
introduction needs to occur as soon as we consider physically
measurable quantities, such as the fields themselves.  } and so we
write, corresponding to~\eqref{eq:4.1},
\begin{equation}
F^{\mu\nu}_{\text{act}} = F^{\mu\nu}_{\text{adv}} + F^{\mu\nu}_{\text{out}}
\label{eq:4.2}
\end{equation}
defining in this way a new field $F^{\mu\nu}_{\text{out}}$. This field
occupies a position symmetrical with our incident field
$F^{\mu\nu}_{\text{in}}$ and so we interpret it as the field of
outgoing radiation leaving the neighbourhood of the electron. The
difference of the two radiation fields
\begin{equation}
F^{\mu\nu}_{\text{rad}} = F^{\mu\nu}_{\text{out}} - F^{\mu\nu}_{\text{in}}
\label{eq:4.3}
\end{equation}
will thus be the radiation field produced by the
electron. From~\eqref{eq:4.1} and~\eqref{eq:4.3} we may express this
radiated field as
\begin{equation}
F^{\mu\nu}_{\text{rad}} = F^{\mu\nu}_{\text{ret}} - F^{\mu\nu}_{\text{adv}}
\label{eq:4.4}
\end{equation}
which shows that $F^{\mu\nu}_{\text{rad}}$ is a function only of the world line of the electron, as
it should be.

Dirac calculates $F^{\mu\nu}_{\text{rad}}$ in the appendix to his
paper, and finds that it is free from singularities and has the value
on the world-line
{
\setlength{\abovedisplayskip}{4pt plus 2pt minus 2pt}%
\setlength{\belowdisplayskip}{4pt plus 2pt minus 2pt}%
\begin{equation}
F^{\mu\nu}_{\text{rad}} = \frac{4e}{3}\left(v^{[\mu}\dot{a}^{\nu]}\right)
\label{eq:4.5}
\end{equation}
We should like to compare definition~\eqref{eq:4.4} with the usual
definition of the radiation field given in the previous chapter. There
we defined the radiation to be given by the field
$F^{\mu\nu}_{\text{ret}}$ at large distances from the charge, and at
correspondingly great times after the proper time which produced the
field. In Dirac's paper he compares the two definitions in the
particular case of a charge moving initially with uniform velocity,
then undergoing a period of acceleration, and then returning to
constant motion again. In this situation it is clear that the right
hand side of~\eqref{eq:4.4} will be the same as
$F^{\mu\nu}_{\text{ret}}$ at large distances from the charge, since
the field $F^{\mu\nu}_{\text{adv}}$ will be the field due to the
motion of the charge in the distant future, i.e.\ the field due to
uniform motion, which we know falls off much faster than the
acceleration field. However if we were to consider an alternative
scenario, for example that of an oscillating charge which we assume to
carry on oscillating into the far distant future, then the argument
above fails because $F^{\mu\nu}_{\text{adv}}$ will make a contribution
to the field far from the charge, and in fact a contribution of the
same magnitude as $F^{\mu\nu}_{\text{ret}}$.

Thus it would appear as if Dirac chose his example explicitly to
justify~\eqref{eq:4.4} for it represents the only circumstance under
which no contradiction with the usual definition of radiation occurs.

It is true that this definition does give a definite value to the radiation
field throughout space-time, unlike the usual definition, and this is an
advantage as it gives meaning to the radiation field close to the electron. Dirac
also concedes that this definition gives meaning to the radiation field before
its time of emission (the advanced contribution), which can have no physical
significance. However he further claims that this is unavoidable if we are
to have the radiation field well defined close to the charge. We know today
that this is false, for Teitelboim \cite{Teitelboim1970} has shown how to construct a meaningful
description of radiation close to a point charge using only retarded fields.

\subsection{The Equations of Motion}
\label{sec:4.2.2}

We now complete our classical theory of the electron by deriving its equation
of motion. To achieve this we apply the principle of conservation of four-
momentum to the system consisting of the charge and the fields described
in the previous section.

We surround the world line of the charge by a thin tube and calculate
the flow of four-momentum across its (3-dimensional) timelike surface, using
the energy-momentum tensor we introduced in Chapter~\ref{chap:3},
\begin{equation*}
T^{\mu\nu} = -\frac{1}{4\pi}\left(F^{\mu\alpha}F_\alpha^{\ \nu}
 + \frac14 F^{\alpha\beta}F_{\alpha\beta} g^{\mu\nu}\right).
\end{equation*}
This is calculated from the field $F^{\mu\nu}_{\text{act}}$. We know
that the total flow of four-momentum into or out of a finite length
of the tube must be the difference between the momentum carried in at
one end of the tube, and the momentum carried out at the other
end. Thus the rate of flow of four-momentum out of the tube must be a
function only of conditions at the ends of the tube, which in
mathematical language means that the rate of flow of four- momentum is
a perfect differential.

At this stage we may ponder whether the result we obtain for the four-
momentum flow will be independent of the shape and size of the
tube. The answer to this question is in the affirmative, as long as
our tube is sufficiently small for the Taylor expansions we use in our
calculations to be valid. That this is so is easily seen by
considering a second tube surrounding the first, whose ends coincide
with the first and calculating the difference between the flow of
four-momentum across this tube, and the flow across the first
tube. The difference is then equal to the integral of $T^{\mu\nu}$
over the surface of the space-time region bounded by the tubes, which
we can convert to an integral of the divergence of $T^{\mu\nu}$
throughout this region by applying Gauss' theorem. Calling $S_1$ the
inner tube, $S_2$ the outer tube and $V$ the volume enclosed between
the two tubes, we have
\begin{align}
\Delta(\text{Four--Momentum Flow})
  &= \int_{S_1} T^{\mu\nu} dS_\nu - \int_{S_2} T^{\mu\nu} dS_\nu \notag\\
  &= \int_V T^{\mu\nu}{}_{,\nu}\, d^4v \notag \\
  &= 0,\label{eq:4.6}
\end{align}
as $T^{\mu\nu}{}_{,\nu}$ is zero in any source-free region of space-time.

The actual calculation involved in getting the four-momentum flow is
very long and complicated and so we shall not give it here. To make life
easier we choose the simplest shape for our tube, a tube that is a sphere of
radius $\epsilon$ in the rest frame of the charge at each value of the proper time $\tau$.
The result of the calculation is that the flow of four-momentum out of the
surface of any length of the tube is given by
\begin{equation}
P^\mu = \int\left[\frac{e^2 a^\mu}{2\epsilon}
   - e F^{\mu\nu}_{\text{in}} v_\nu
   - \frac{2 e^2}{3}\left(\dot{a}^\mu - a^2 v^\mu\right)\right] d\tau
\label{eq:4.7}
\end{equation}
where terms that vanish with $\epsilon$ have been neglected, and the integration is
taken over the length of the tube in question. Neglecting terms of order
$\epsilon$ and greater is tantamount to assuming that we are taking the limit as
$\epsilon$ tends to zero, i.e.\ we are considering the limiting case of a tube of zero
radius surrounding the world line. The importance of this observation will
be apparent when we come to discuss the equation of motion.

We now apply the aforementioned condition that the integrand above
must be a perfect differential, and set accordingly,
\begin{equation}
\left[\frac{e^2 a^\mu}{2\epsilon}
   - e F^{\mu\nu}_{\text{in}} v_\nu
   - \frac{2 e^2}{3}\left(\dot{a}^\mu - a^2 v^\mu\right)\right]
   = \dot{B}^\mu
\label{eq:4.8}
\end{equation}

This is as far as we can get with the law of conservation of
four-momentum, and so to fix our equation of motion for the charged
particle we must make some further assumptions about the four-vector
$B^\mu$. By contracting \eqref{eq:4.8} with the four-velocity $v_\mu$,
and applying the identities \eqref{eq:2.21}, \eqref{eq:2.20} and
\eqref{eq:2.9}, we find
\begin{equation}
\dot{B}^\mu v_\mu = 0
\label{eq:4.9}
\end{equation}
The simplest such $B^\mu$ satisfying \eqref{eq:4.9} is
\begin{equation}
B^\mu = k v^\mu
\label{eq:4.10}
\end{equation}
where $k$ is any constant. There are, of course, other possible expressions for
$B^\mu$, the next simplest one being
\begin{equation}
B^\mu = k\left(a^4 v^\mu - 4(a^\nu \dot{a}_\nu)a^\mu\right)
\label{eq:4.11}
\end{equation}
but since they are all far more complicated than \eqref{eq:4.10} we appeal to simplicity
and assume \eqref{eq:4.10}. Substituting \eqref{eq:4.10} into the right hand side of \eqref{eq:4.8} we have
the equation
\begin{equation}
\left(\frac{e^2}{2\epsilon} - k\right) a^\mu
  - \frac{2 e^2}{3}(\dot{a}^\mu - a^2 v^\mu)
  = e F^{\mu\nu}_{\text{in}} v_\nu
\label{eq:4.12}
\end{equation}

\subsection{Mass Renormalisation}
\label{sec:4.2.3}

In order to connect \eqref{eq:4.12} with the motion of a point charge, we need to
identify in \eqref{eq:4.12} a term associated with the mass of the charge. Since the
left hand side of \eqref{eq:4.12} represents the reaction of the world-line of the charge
when acted upon by an external field $F^{\mu\nu}_{\text{in}}$, we suppose that the coefficient
in front of the acceleration $a^\mu$ is in fact the usual rest mass of the particle.
Thus we identify
\begin{equation}
\frac{e^2}{2\epsilon} - k = m
\label{eq:4.13}
\end{equation}
and so our equation of motion becomes
\begin{equation}
m a^\mu - \frac{2 e^2}{3}(\dot{a}^\mu - a^2 v^\mu)
  = e F^{\mu\nu}_{\text{in}} v_\nu
\label{eq:4.14}
\end{equation}
This is the equation of motion we set out to find, and it is known as
the \emph{Lorentz--Dirac Equation}.

Now some confusion arises. What are we to make of the $1/\epsilon$
term? As already mentioned, this whole derivation is only strictly
accurate in the limit as $\epsilon$ tends to zero, and so for $m$ to
be the observed rest mass of the point charge (e.g.\ for the electron
$9.1\times 10^{-28}$ grams), we see from \eqref{eq:4.13} that $k$ must be an
infinite quantity. Now, by Eqs.~\eqref{eq:4.8} and \eqref{eq:4.7}, we have that the flow
of four-momentum out of the tube surrounding the world-line between
proper time $\tau_1$ and proper time $\tau_2$ ($\tau_2 > \tau_1$) is
$B^\mu(\tau_2) - B^\mu(\tau_1)$. Thus $B^\mu(\tau)$ should be
interpreted as minus the four-momentum residing within the tube at
proper time $\tau$ and so, from \eqref{eq:4.10} and \eqref{eq:4.13}, the energy ($P^0 =
-B^0$) within the tube is negative and must tend to $-\infty$ as
$\epsilon$ tends to zero. As Dirac puts it, this negative energy
within the tube is needed to cancel the large positive energy of the
Coulomb field lying outside the tube, to keep the total energy down to
the rest-mass of the point-charge. Thus we are forced to the
conclusion that a point charge consists of an infinite, negative mass
at its center, such that when it is subtracted from the infinite
positive energy of the surrounding Coulomb field we are left with a
finite result whose value is just $m$. This is called
mass-renormalisation.

Mass renormalisation can be justified on physical grounds by imagining
what would happen if we tried to accelerate a point charged particle that
had a finite mass center $m_o$ (positive or negative, it makes no difference).
For definiteness let's assume that we're initially observing the particle from
its rest-frame, and that the particle has been in uniform motion for its entire
history. Its four-momentum will be
\begin{equation*}
P^\mu = (\infty + m_o, 0,0,0) = m v^\mu.
\end{equation*}
The infinite term represents the energy in the Coulomb field of the charge\footnote{
That the Coulomb field of a point particle contains an infinite amount of energy is
easily seen by integrating the energy density of the field ($E^2$) over all space in the rest-
frame of the charge. Because $E\propto 1/R^2$ where $R$ is the spatial distance from the charge,
the integral will diverge.
}
If we try to accelerate the charge by some means (perhaps by applying an
external field), then we must apply a four-force equal to the rate of change
of its four-momentum, i.e.
\begin{equation}
F^\mu = m a^\mu
\label{eq:4.15}
\end{equation}
}
where we are neglecting the radiation terms in \eqref{eq:4.14}. But in this case
$m = \infty + m_o = \infty$ since $m_o > -\infty$ and so if $a^\mu \neq 0$ then we must apply an
infinite force to move the charge. This infinite force is needed to transport
the infinite energy in the Coulomb field of the charge. Thus under practical
circumstances we would not be able to move any point charge, which is
clearly preposterous since the electron is such a charge (as far as we can
determine) and yet it resists our attempts to move it, not with an infinite
mass, but with a mass of $9.1\times 10^{-28}$ grams.

Looking at the individual terms in \eqref{eq:4.14} we see that the right hand
side is the four-force exerted by the incident field upon the
charge. The first term on the left hand side, $m a^\mu$ we have
already identified as the rate of change of four-momentum of the
Coulomb field plus the central mass, and as such represents changes in
the four-momentum `carried along' by the charge. The third term,
$\frac23a^2v^\mu$, from Chapter~\ref{chap:3}, is the rate at which the
charge is radiating away four-momentum in its field. The second term
is not so easily interpreted. We know from Chapter~\ref{chap:3} that
it has nothing to do with radiation and so the only acceptable picture
we can have for this term is that it represents a kind of reversible
emission of radiation that never gets very far from the charge.

It is interesting to compare our first attempt at an equation of
motion, \eqref{eq:3.29} with \eqref{eq:4.14}. The latter equation is
identical to the former except for the $\frac23\dot{a}^\mu$ term. It
is this term which fixes up the problem we had with \eqref{eq:3.29}
when contracting with $v^\mu$. Instead of getting $a^2 = 0$, as we did
before, \eqref{eq:4.14} yields the identity $a^2 - a^2 = 0$. In fact
the term $\frac23\dot{a}^\mu$ is the simplest modification we could
make to \eqref{eq:3.29} to rectify the contraction problem.

Although the Lorentz–Dirac equation is well accepted today as the classical
equation of motion for a point charge, the derivation we have presented
here is fraught with interpretational difficulties. In the next section we give
an alternative derivation which in our opinion is far easier to understand
and more physically acceptable than Dirac's procedure.

\section{Teitelboim's Procedure}
\label{sec:4.3}

The main advantage of Dirac's definition of the radiation field
\eqref{eq:4.5} is that we can use it to define radiation all the way
up to a point charge, not just at infinity as we did in
Chapter~\ref{chap:3}. This is a desirable situation as radiation
corresponds to the emission of something (photons) from a charge and
so we would expect such an emission to be well defined at the world
line of the charge and not just at infinity. Dirac claimed that to
have such a definition of radiation we are forced to introduce
advanced fields, and hence saddle ourselves with all interpretational
problems associated with such fields. In fact this claim is incorrect,
and to see this we can look back at our discussion of radiation from
Chapter~\ref{chap:3}. There we calculated the four-momentum radiated
away at infinity by integrating the energy-momentum tensor over the
surface of a sphere of radius $R$ in the rest-frame of the charge with
origin at the charge. We then took the limit as $R$ diverged to
infinity, and in this way saw that only the far field contributed to
radiation, and in fact the far part of the energy-momentum tensor
$T^{\mu\nu}_{II}$ \eqref{eq:3.19} was such that when integrated over
the surface of the sphere the result was independent of $R$. This
gives a very big clue on how to define radiation all the way up to the
charge without introducing advanced fields. We simply take the
radiated four-momentum at any distance, $d$, from the charge (in its
rest frame) to be the integral of $T^{\mu\nu}_{II}$ over the surface
of a sphere centered on the charge, and of radius $d$.

This definition of radiation via only retarded fields is the central
point of Teitelboim's paper \cite{Teitelboim1970}. In addition to defining
radiated four-momentum through a splitting of the electromagnetic
field tensor into a near and far part, he also defines the bound
four-momentum in the field in a similar way. We will discuss in full
all these features of Teitelboim's treatment.

\subsection{Splitting of the energy-momentum tensor}
\label{sec:4.3.1}

We split the retarded field tensor for a point charge into the near
and far fields as in Chapter~\ref{chap:3}. Reiterating those results,
we have
{
\setlength{\abovedisplayskip}{3pt plus 2pt minus 2pt}%
\setlength{\belowdisplayskip}{3pt plus 2pt minus 2pt}%
\begin{equation}
F^{\mu\nu} = F^{\mu\nu}_I + F^{\mu\nu}_{II}
\label{eq:4.16}
\end{equation}
where, as before,
\begin{equation}
F^{\mu\nu}_I = \frac{e}{\rho^3} v^{[\mu} r^{\nu]}
\label{eq:4.17}
\end{equation}
is the near field, and
\begin{equation}
F^{\mu\nu}_{II} = \frac{e}{\rho^2}\left(a_r v^{[\mu} r^{\nu]} + a^{[\mu} r^{\nu]}\right)
\label{eq:4.18}
\end{equation}
is the far or radiation field.

Substituting \eqref{eq:4.17} and \eqref{eq:4.18} into the expression
for the energy momentum tensor, \eqref{eq:3.5}
\begin{equation*}
T^{\mu\nu} = -\frac{1}{4\pi}\left(F^{\mu\alpha}F_{\alpha}{}^{\nu} + \frac{1}{4}F^{\alpha\beta}F_{\alpha\beta}g^{\mu\nu}\right).
\end{equation*}
we get
\begin{equation}
T^{\mu\nu} = T^{\mu\nu}_I + T^{\mu\nu}_{I,II} + T^{\mu\nu}_{II},
\label{eq:4.19}
\end{equation}
where $T^{\mu\nu}_I$ and $T^{\mu\nu}_{II}$ are the tensors obtained
when \eqref{eq:3.5} is evaluated with the fields $F^{\mu\nu}_I$ and
$F^{\mu\nu}_{II}$ respectively, and $T^{\mu\nu}_{I,II}$ is the
combination of cross terms containing both $F^{\mu\nu}_I$ and
$F^{\mu\nu}_{II}$. Explicitly, these tensors are
\begin{align}
T^{\mu\nu}_I
&= -\frac{e^2}{4\pi\rho^4}\left(\frac{r^\mu r^\nu}{\rho^2} - \frac{v^{(\mu}r^{\nu)}}{\rho} - \frac{1}{2}g^{\mu\nu}\right),
\label{eq:4.20}\\
T^{\mu\nu}_{I,II}
&= -\frac{e^2}{4\pi\rho^3}\left(2a_r\frac{r^\mu r^\nu}{\rho^2} - \frac{1}{\rho}\left(a_r v^{(\mu}r^{\nu)} + a^{(\mu}r^{\nu)}\right)\right),
\label{eq:4.21}\\
T^{\mu\nu}_{II}
&= -\frac{e^2 r^\mu r^\nu}{4\pi\rho^4}\left(a_r^2-a^2\right),
\label{eq:4.22}
\end{align}
with
\begin{equation*}
a^{(\mu}b^{\nu)} = a^\mu b^\nu + a^\nu b^\mu.
\end{equation*}

It is useful to know the divergences of each term above. We know from
Chapter~\ref{chap:3} that $T^{\mu\nu}$ satisfies
\begin{equation}
T^{\mu\nu}{}_{,\nu} = 0,
\label{eq:4.23}
\end{equation}
off the world line of the charge. We use the retarded differentiation
techniques developed in Chapter~\ref{chap:2} to derive the
corresponding formulae for $T^{\mu\nu}_I$, $T^{\mu\nu}_{II}$ and
$T^{\mu\nu}_{I,II}$:
\begin{align}
T^{\mu\nu}_{I,\nu} &= \frac{e^2 a_r}{2\pi\rho^5}\,r^\mu
\label{eq:4.24}\\
T^{\mu\nu}_{II,\nu} &= 0
\label{eq:4.25}\\
T^{\mu\nu}_{I,II,\nu} &= -\frac{e^2 a_r}{2\pi\rho^5}\,r^\mu
\label{eq:4.26}
\end{align}
}
These results are only valid off the world line of the charge (we
can't differentiate $F^{\mu\nu}$ on the world-line as it is singular
there). Equations \eqref{eq:4.24} and \eqref{eq:4.26} suggest that we
define a new tensor,
\begin{equation}
T^{\mu\nu}_s = T^{\mu\nu}_I + T^{\mu\nu}_{I,II}
\label{eq:4.27}
\end{equation}
so that
\begin{equation}
T^{\mu\nu} = T^{\mu\nu}_s + T^{\mu\nu}_{II}
\label{eq:4.28}
\end{equation}
with
\begin{align}
T^{\mu\nu}_{s,\nu} &= 0,
\label{eq:4.29}\\
T^{\mu\nu}_{II,\nu} &= 0.
\label{eq:4.30}
\end{align}

Equations \eqref{eq:4.29} and \eqref{eq:4.30} are again only valid off
the world line. We note that the tensor $T^{\mu\nu}_{II}$ is related
only to the far field of the charge, while $T^{\mu\nu}_s$ is the
combination of the near field and the interference between the near
and far fields.

Thus we arrive at the conclusion that a splitting of the Maxwell
electromagnetic field tensor into a near and a far part induces a
splitting of the full energy-momentum tensor into two parts, both of
which are conserved off the world line of the particle and so are
energy-momentum tensors in their own right.

\subsection{Definition of Four-Momentum}
\label{sec:4.3.2}

Having discovered that the electromagnetic energy-momentum tensor for
a point charge splits into two separately conserved parts we would now
like to find the field four-momentum corresponding to each part.
\clearpage
\begin{figure}[t]
  \vspace*{-1.8cm}
  \centering
  \includegraphics[width=\textwidth]{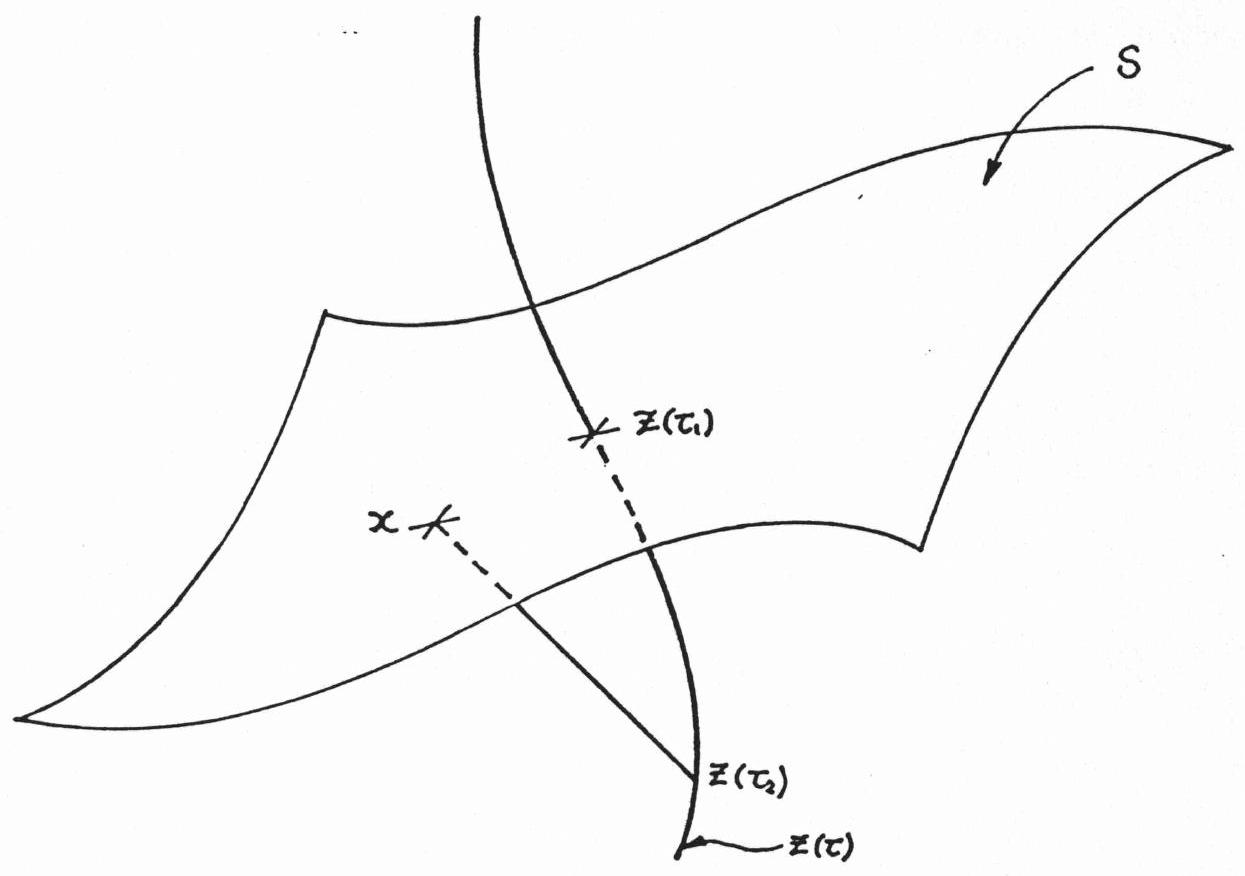}
  \caption{Spacelike surface intersecting world-line. $(x^\mu -
    z^\mu(\tau_2))(x_\mu - z_\mu(\tau_2)) = 0$.}
  \label{fig:4.1}
\end{figure}

Following \S\ref{sec:3.2} we have the usual definition of the
four-momentum of the field:
\begin{equation}
  P^\mu = \int_{S} T^{\mu\nu} dS_\nu
  \label{eq:4.31}
\end{equation}
where $S$ is any infinite space-like surface intersecting the
world-line of the charge, and $dS_\nu$ is the future-pointing
surface-element.

$P^\mu$ defined by \eqref{eq:4.31} will only be conserved for
electromagnetic fields in a source free region of space-time, for as
we saw in Chapter~\ref{chap:3}, any currents will effectively `pump'
four-momentum into the field. How then are we to make this definition
apply in our case when we do indeed have a source present in the
region of integration? The answer to this question lies in a
consideration of the geometry of the problem. With reference to
Figure~\ref{fig:4.1} we can see that if our space-like surface of
integration intersects the world-line of the charge at a point
$z(\tau_1)$, then the field at any point on that surface, say $x$,
will be a function of the corresponding retarded world-line point,
$z(\tau_2)$. Because of the space-like nature of $S$ we have that
$\tau_2 < \tau_1$, and thus the integral in \eqref{eq:4.31} will be
completely determined by the world-line of the charge prior to $\tau =
\tau_1$. This is true regardless of how the surface $S$ is chosen, and so in
order that $P^\mu$ be unambiguously defined, we would hope that the
right hand side of \eqref{eq:4.31} is independent of $S$, and hence
the following definition would be natural:
\begin{equation}
  P^\mu(\tau) = \int_{S} T^{\mu\nu} dS_\nu.
  \label{eq:4.32}
\end{equation}
where now $S$ is any space-like surface intersecting the world-line of
the charge at $z(\tau)$. In particular, invariance of definition
\eqref{eq:4.32} under \textit{tilting} of $S$ will make it a true
four-vector.

Unfortunately it is not true that this integral is invariant under
transformations in the surface $S$. However, all is not lost, for if
we replace $T^{\mu\nu}$ by $T^{\mu\nu}_{II}$ in the above definition
then this modified integral is invariant under changes in
$S$. Accordingly we define
\begin{equation}
  P^\mu_{II}(\tau) = \int_{S} T^{\mu\nu}_{II} dS_\nu.
  \label{eq:4.33}
\end{equation}

Teitelboim proves the invariance of \eqref{eq:4.33} under changes in
$S$ by dividing the $S$ up into strips and considering each strip
separately. We give a somewhat more immediate proof than Teitelboim's
in what follows.

\begin{figure}[t]
  \vspace*{-1.8cm}
  \centering
  \includegraphics[width=\textwidth]{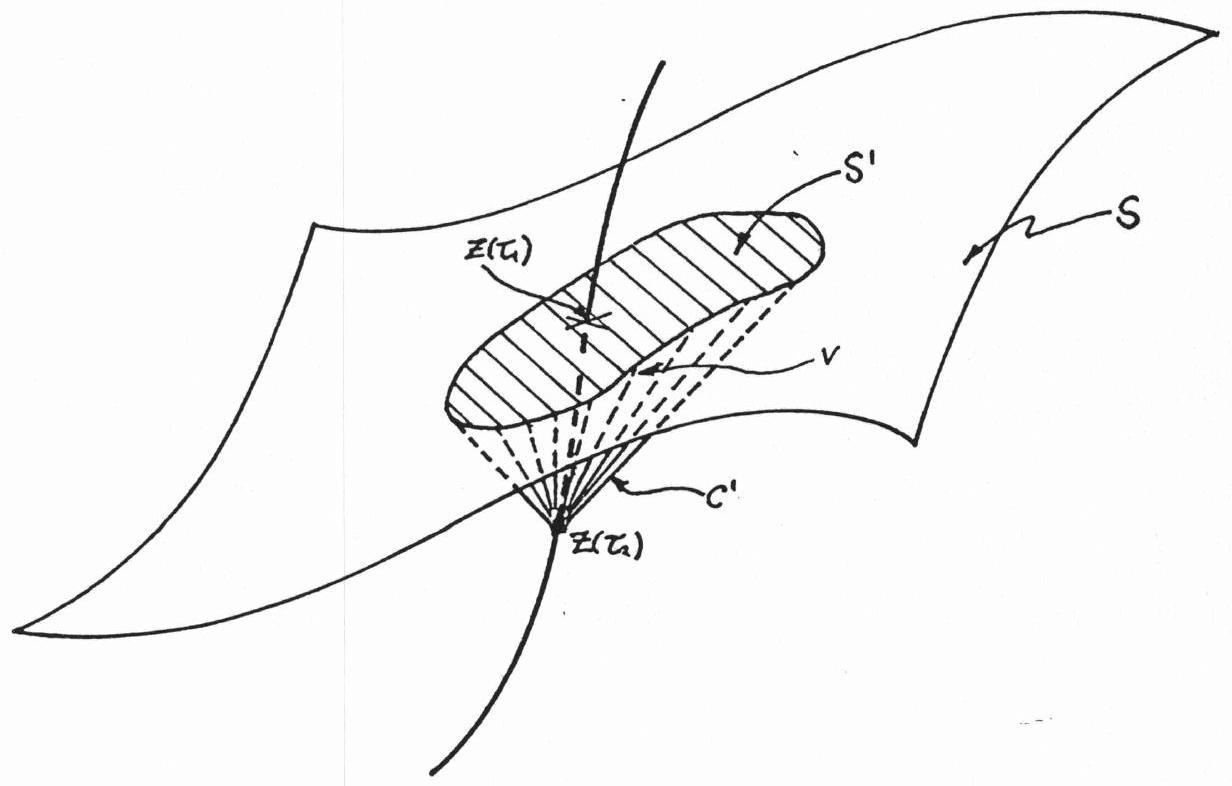}
  \caption{Intersection of light-cone from $z(\tau_2)$ with spacelike surface $S$}
  \label{fig:4.2}
\end{figure}

We first assume that the integral in \eqref{eq:4.33} is finite, and
that integration over any subset of $S$ is finite. This assumption
will be justified later. With reference to Figure~\ref{fig:4.2} we
consider a region of space-time bounded by the surface $S$ which
intersects the world-line at $z(\tau_1)$, and the future light-cone
from the world-line at some point $z(\tau_2)$ where $\tau_2 < \tau_1$.

{
\setlength{\abovedisplayskip}{4pt plus 2pt minus 2pt}%
\setlength{\belowdisplayskip}{4pt plus 2pt minus 2pt}%
Denoting the enclosed volume by $V$, and the light-cone surface by
$C$, we have by Gauss's theorem
\begin{equation}
  \int_{V} T^{\mu\nu}_{II,\nu}\, d^4 v
  =
  \int_{S'} T^{\mu\nu}_{II}\, dS_\nu
  +
  \int_{C'} T^{\mu\nu}_{II}\, dC_\nu.
  \label{eq:4.34}
\end{equation}
where $S'$ and $C'$ are the parts of $S$ and $C$ forming the boundary
of $V$. From \eqref{eq:4.22} we have
\[
  T^{\mu\nu}_{II} \propto r^\mu r^\nu
\]
and as $r^\nu$ is a null vector we get
\[
  T^{\mu\nu}_{II}\, dC_\nu = 0
\]
since $dC_\nu$ is a light cone surface element and hence is null
itself. Thus \eqref{eq:4.34} becomes
\begin{equation}
  \int_{V} T^{\mu\nu}_{II,\nu}\, d^4 v
  =
  \int_{S'} T^{\mu\nu}_{II}\, dS_\nu.
  \label{eq:4.35}
\end{equation}
The integrand of the left-hand side of \eqref{eq:4.35} is only
non-zero on the world-line, by \eqref{eq:4.30}. Thus the left-hand
side is dependent only on the nature of the world-line between
$\tau_2$ and $\tau_1$, and not on anything else to do with the volume
of integration. Hence, the only feature of $S$ that affects this
integral is where it intersects the world-line. If we now let $\tau_2$
tend to $-\infty$ we find that $P^\mu_{II}(\tau)$ given by
\eqref{eq:4.33} is a function only of $\tau$, and is independent of
the surface $S$, as we wished to prove.

That $P^\mu_{II}(\tau)$ is independent of $S$ justifies calling it the
\textit{emitted four-momentum}. It shows that the rest-frame of the
charge does not occupy a privileged role when it comes to calculating
this momentum, further reinforcing the fact that this part of the
field, once emitted, detaches itself from the charge.

Now we return to the original problem of how to define the
four-momentum for the full field, and not just for the field
associated with $T^{\mu\nu}_{II}$. Substituting \eqref{eq:4.28} into
\eqref{eq:4.32} we have
\begin{equation}
  P^\mu(\tau) = \int_{S} T^{\mu\nu}_{s}\, dS_\nu + P^\mu_{II}(\tau).
  \label{eq:4.36}
\end{equation}
Since $P^\mu_{II}$ appears to be well behaved, any problems associated
with this definition must stem from the energy-momentum tensor
$T^{\mu\nu}_{s}$. Certainly our previous proof of the
surface-independence of $P^\mu_{II}$ will not apply to $P_s^\mu$ defined
in a similar way. This is easily seen from the fact that
$T^{\mu\nu}_{s}$ is not proportional to $r^\mu r^\nu$, a fact that was
pivotal in our proof. Additionally, we know that the four-momentum in
the Coulomb field of a point charge is infinite, and the density of
this appears in $T^{\mu\nu}_{s}$.

It would thus seem inappropriate to associate a vector of
four-momentum with $T^{\mu\nu}_{s}$, for it appears that such a vector
will not satisfy even the most basic requirements of a vector of
four-momentum. However Teitelboim proposes that it is exactly these
failings which indicate that the four-momentum associated with
$T^{\mu\nu}_{s}$ is special, and in fact he argues that the dependence
of this four-momentum on the surface of integration implies that it is
somehow `bound' to the charge. If this is so then we would expect the
rest-frame of the charge to play a privileged role in the definition
of four-momentum corresponding to this part of the energy-momentum
tensor. We thus define the bound four-momentum present at proper time
$\tau$ by
\begin{equation}
  P^\mu_{s}(\tau) = -\int_{\sigma(\tau)} T^{\mu\nu}_{s}\, v_\nu(\tau)\, d^3\sigma.
  \label{eq:4.37}
\end{equation}
where $\sigma(\tau)$ is the spacelike surface defined by coordinates $x^\mu$ satisfying
\begin{equation}
  v^\mu(\tau)\, (x_\mu - z_\mu(\tau)) = 0.
  \label{eq:4.38}
\end{equation}
}
The measure element on $\sigma(\tau)$ is related to the volume element
in the rest-frame by $d^3\sigma = d^3x$.

We now have definitions for both the bound and emitted four-momentum
in the field of a point charge. To derive the equation of motion we
need to calculate explicitly these quantities and in particular their
rates of change. This is the subject of the following two sections.

\subsection{Calculation of Emitted Four-Momentum}
\label{sec:4.3.3}

\begin{figure}[t]
  \vspace*{-1.8cm}
  \centering
  \includegraphics[width=0.65\textwidth]{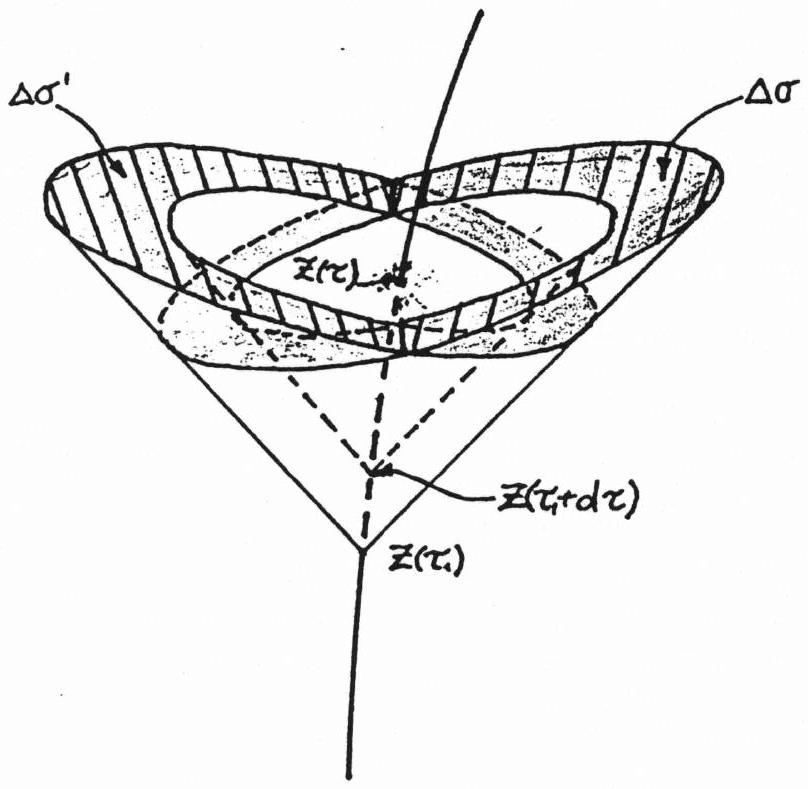}
  \caption{Intersection of future light-cone from $z(\tau_1)$ and
    $z(\tau_1 + d\tau)$ with two spacelike surfaces cutting the
    world-line at $z(\tau)$.}
  \label{fig:4.3}
\end{figure}

To compute $P^\mu_{II}(\tau)$ we need to evaluate the integral in
\eqref{eq:4.33}. We begin by choosing an arbitrary surface, $S$, over
which we shall perform the integration. With reference to
Figure~\ref{fig:4.3}, we consider the contribution to the integral by
the region of world-line between proper-time $\tau_1$, and an
infinitesimal time later $\tau_1 + d\tau$. If we denote by
$\Delta\sigma$ the region of $S$ lying between the two future light
cones from $z(\tau_1)$ and $z(\tau_1 + d\tau)$, then we have
\begin{equation}
\Delta P^\mu_{II}(\tau) = \int_{\Delta\sigma} T^{\mu\nu}_{II}\, dS_\nu.
\label{eq:4.39}
\end{equation}

Now consider an alternative surface $S'$, also intersecting the
world-line at $z(\tau)$. This surface will also have a region lying
between the two light-cones, and we denote this by
$\Delta\sigma'$. Corresponding to this new region we write
\begin{equation}
\Delta P^{\prime\mu}_{II}(\tau) = \int_{\Delta\sigma'} T^{\mu\nu}_{II}\, dS'_\nu.
\label{eq:4.40}
\end{equation}
Applying Gauss' theorem to the volume bounded by the two light-cones,
$\Delta\sigma$ and $\Delta\sigma'$, and recalling that the divergence
of $T^{\mu\nu}_{II}$ is zero and that the flux of $T^{\mu\nu}_{II}$
across the light cones is zero, we arrive at the result
\begin{equation}
\Delta P^\mu_{II}(\tau) = \Delta P^{\prime\mu}_{II}(\tau).
\label{eq:4.41}
\end{equation}

{
\setlength{\abovedisplayskip}{4pt plus 2pt minus 2pt}%
\setlength{\belowdisplayskip}{4pt plus 2pt minus 2pt}%
Using equation \eqref{eq:4.41}, we choose $S'$ to be the three-space
of the rest-frame of $z^\mu(\tau_1)$. We have already considered
calculations of four-momentum flux in this frame in \S\ref{sec:3.3.1},
and so borrowing the notation of that section, the integral
\eqref{eq:4.39} takes the form
\begin{equation}
\int_{\Delta\sigma} T^{\mu\nu}_{II}\, dS_\nu = d\tau \int_{sphere} T^{\mu 0}_{II}\, dA
\label{eq:4.42}
\end{equation}
Recalling \eqref{eq:3.20} to \eqref{eq:3.22} and imitating the
calculations following those equations, we obtain
\begin{equation}
\Delta P^\mu_{II}(\tau) = d\tau \frac{2e^2}{3}a^2(\tau_1)(1,0,0,0)
\label{eq:4.43}
\end{equation}
Thus in an arbitrary frame we have
\begin{equation}
\Delta P^\mu_{II}(\tau) = d\tau \frac{2e^2}{3}a^2(\tau_1)v^\mu(\tau_1),
\label{eq:4.44}
\end{equation}
or
\begin{equation}
\left.\frac{dP^\mu_{II}(\tau)}{d\tau}\right|_{\tau=\tau_1} = \frac{2e^2}{3}a^2(\tau_1)v^\mu(\tau_1).
\label{eq:4.45}
\end{equation}
Note that the radiation rate \eqref{eq:4.45} is consistent with
\eqref{eq:3.27}.

To obtain the total momentum of type $II$ present at time $\tau$ we
integrate \eqref{eq:4.45} along the whole of the past world-line, i.e.
\begin{equation}
P^\mu_{II}(\tau) = \frac{2e^2}{3}\int_{-\infty}^{\tau} a^2(\tau')v^\mu(\tau')d\tau'.
\label{eq:4.46}
\end{equation}
}

It is important to emphasize the dependence of \eqref{eq:4.46} on the
whole past history of the world-line. This is a consequence of the
fact that $P^\mu_{II}$ corresponds to the emitted four-momentum in the
field.

\begin{figure}[t]
  \vspace*{-2.2cm}
  \centering
  \includegraphics[width=0.75\textwidth]{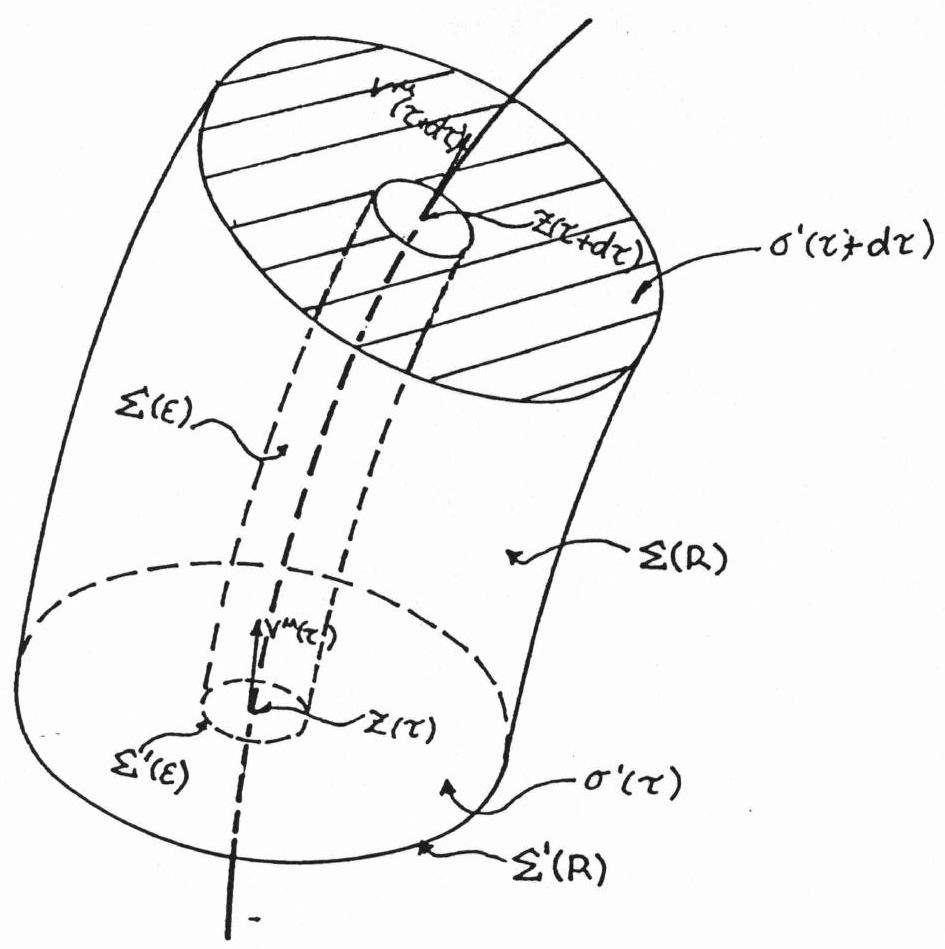}
  \caption{Evaluation of the rate of change of bound four-momentum, $P^\mu_s$.}
  \label{fig:4.4}
\end{figure}

We now turn our attention to the remaining part of the electromagnetic four-momentum
$P^\mu_s(\tau)$.

\subsection{Calculation of Bound Four-Momentum}
\label{sec:4.3.4}

To find $P^\mu_s(\tau)$ as defined by equation \eqref{eq:4.37} we first evaluate its rate of
change per unit of proper-time and then integrate the resulting expression.

To compute the rate of change of the bound four-momentum we surround the
world-line of the charge by two timelike tubes, similar to the tube used
in the Dirac procedure. The first tube we take to be of vanishingly small
radius $\epsilon$, and the second to be of divergently large radius $R$.
The surfaces of these tubes we denote by $\Sigma(\epsilon)$ and $\Sigma(R)$
respectively, and as with the Dirac procedure they are spheres of radii
$\epsilon$ and $R$ in the rest-frame of the charge at each proper time $\tau$.
The situation is pictured in Figure~\ref{fig:4.4}.

Mathematically $\Sigma(\epsilon)$ is defined by coordinates $x^\mu$ satisfying the equations
\begin{align}
  (x-z(s))^2 &= \epsilon^2
  \label{eq:4.47}\\
  v^\mu(s)\,(x_\mu - z_\mu(s)) &= 0.
  \label{eq:4.48}
\end{align}
(We are using $s$ as a `variable' proper-time parameter to avoid confusion
with the `fixed' proper-time $\tau$ appearing in $P^\mu_I(\tau)$.)
To find the surface measure element of $\Sigma(\epsilon)$, consider a
variation of a point $x^\mu$ on the surface to the point $x^\mu + dx^\mu$,
also on the surface. Let the corresponding variation in $s$ be $ds$.
Then we have, from \eqref{eq:4.47} and \eqref{eq:4.48},
\begin{align*}
  (x^\mu - z^\mu(s))\,(dx_\mu - v_\mu(s)ds) &= 0,\\
  v^\mu(s)(dx_\mu - v_\mu(s)ds) + a^\mu(s)ds(x_\mu - z_\mu(s)) &= 0,
\end{align*}
which reduce, with the help of the identities \eqref{eq:2.20} and \eqref{eq:4.47}, to
\begin{align*}
  (x^\mu - z^\mu(s))\,dx_\mu &= 0\\
  v^\mu(s)\,dx_\mu &= -(1 + a^\mu(s)(x_\mu - z_\mu(s)))\,ds.
\end{align*}

Making the abbreviation, $n^\mu = (x^\mu - z^\mu(s))/\epsilon$, and writing $a_n$ for
$a^\mu(s)n_\mu$, the above equations can be rendered in the form
\begin{align}
  n^\mu dx_\mu &= 0
  \label{eq:4.49}\\
  v^\mu(s)dx_\mu &= -(1+\epsilon a_n)ds.
  \label{eq:4.50}
\end{align}

Equation \eqref{eq:4.49} shows us that $n^\mu$ is the unit normal to
$\Sigma(\epsilon)$ ($n^\mu n_\mu = \epsilon^2/\epsilon^2 = 1$), as we
would expect. Equation \eqref{eq:4.50} indicates that the part of the
surface element parallel to $v^\mu$ is $(1+\epsilon a_n)v^\mu ds$. Now
the full 3-surface measure element of $\Sigma(\epsilon)$ is equal to
the two-dimensional measure element of a section of the surface by a
three-dimensional plane orthogonal to $v^\mu(s)$, multiplied by the
surface element parallel to $v^\mu$. We already know that the
two-dimensional surface-element given by such a section is just
$\epsilon^2d\Omega$ where $d\Omega$ is the solid angle element
subtended by the space-part of $n^\mu$ in the rest-frame of the charge
at proper-time $s$, and so the full 3-surface measure element is given
by
\begin{equation}
  d^3\sigma \;=\; \epsilon^2(1+\epsilon a_n)\,d\Omega\,ds.
  \label{eq:4.51}
\end{equation}

To obtain the corresponding results for $\Sigma(R)$ it suffices to replace $\epsilon$
by $R$ in the preceding expressions.

To proceed with our calculation, we now apply Gauss' theorem to the
region of space-time bounded by the two tubes and the two surfaces
$\sigma(\tau)$ and $\sigma(\tau + d\tau)$, where $d\tau$ is
infinitesimal (\textit{cf.}~\eqref{eq:4.38} and
Figure~\ref{fig:4.4}). As $T^{\mu\nu}_s$ is divergence free in such a
region (the world line has been explicitly excluded), and noting that
the normal to $\sigma(\tau)$ is $v^\mu(\tau)$, we find
\begin{align*}
  P^\mu_s(\tau + d\tau) - P^\mu_s(\tau)
  &= -\int_{\sigma(\tau+d\tau)} T^{\mu\nu}_s\,v_\nu(\tau+d\tau)\,d^3\sigma
     + \int_{\sigma(\tau)} T^{\mu\nu}_s\,v_\nu(\tau)\,d^3\sigma\\
  &= \lim_{\epsilon\to0}\int_{\Sigma(\epsilon)} T^{\mu\nu}_s\,n_\nu\,d^3\sigma
     - \lim_{R\to\infty}\int_{\Sigma(R)} T^{\mu\nu}_s\,n_\nu\,d^3\sigma
\end{align*}
As $d\tau$ is infinitesimal, we can write this as
\begin{align*}
  P^\mu_s(\tau + d\tau) - P^\mu_s(\tau)
  = d\tau\Bigg(&\lim_{\epsilon\to0}\int_{\Sigma'(\epsilon)} T^{\mu\nu}_s\,n_\nu\,
      \epsilon^2(1+\epsilon a_n)\,d\Omega\\
  &\qquad\qquad - \lim_{R\to\infty}\int_{\Sigma'(R)} T^{\mu\nu}_s\,n_\nu\,
      R^2(1+Ra_n)\,d\Omega\Bigg),
\end{align*}
where $\Sigma'(\epsilon)$ and $\Sigma'(R)$ are the (2-surface)
intersections of $\Sigma(\epsilon)$ and $\Sigma(R)$ with
$\sigma(\tau)$, and $a$, $n$ are evaluated at proper-time
$\tau$. Dividing by $d\tau$, we then have
\begin{equation}
  \frac{dP^\mu_s}{d\tau}
  = \lim_{\epsilon\to0}\int_{\Sigma'(\epsilon)} T^{\mu\nu}_s\,n_\nu\,\epsilon^2(1+\epsilon a_n)\,d\Omega
    - \lim_{R\to\infty}\int_{\Sigma'(R)} T^{\mu\nu}_s\,n_\nu\,R^2(1+Ra_n)\,d\Omega.
  \label{eq:4.52}
\end{equation}

Teitelboim now proves that under the condition of a straight world-line
(i.e.\ uniform motion) in the remote past the integral over $\Sigma'(R)$ in
\eqref{eq:4.52} vanishes as $R$ tends to infinity. We give a slightly quicker
proof than Teitelboim's.

From \eqref{eq:4.20} and \eqref{eq:4.21} we see that $T^{\mu\nu}_s$ is
of the form $A^{\mu\nu}/\rho^4 + B^{\mu\nu}/\rho^3$ with $A^{\mu\nu}$
and $B^{\mu\nu}$ finite in the limit $\rho \to \infty$. In the
integration $\rho$ will be the spatial distance from any point on
$\Sigma'(R)$ to the corresponding retarded world-line point,
\textit{as viewed from the rest-frame of the charge at that point}
(\textit{cf.}\ Chapter~\ref{chap:2}). We would like to find some
relationship now between $\rho$ and $R$, in particular a bound on
their ratio.

Clearly if we allow our charge to have a world line which approaches
asymptotically close to the speed of light as $\tau \to -\infty$ then
we can have a situation where $R/\rho \to 0$ as $R \to
\infty$. However such a scenario is unphysical as it requires our
(massive) charge to be travelling at the speed of light at $\tau =
-\infty$. So we assume there is an upper bound to the velocity of the
charge, \textit{as viewed from the rest-frame of the charge at any
  proper-time $\tau$}.  It is necessary to include this proviso as we
can jack the velocity of the charge up as far as we like if we allow
ourselves to view it from an arbitrary frame. Accordingly, we write
\begin{equation}
  |v^\mu(\tau)| < v_{\max}
  \label{eq:4.53}
\end{equation}
for $\mu = 0,1,2,3$ and this equation is true for all $\tau$ in any rest-frame of the
charge. We denote the maximum speed corresponding to $v_{\max}$ by $\mu$, where
$\mu < 1$.

\begin{figure}[!t]
  \vspace*{-2.5cm}
  \includegraphics[width=0.95\textwidth]{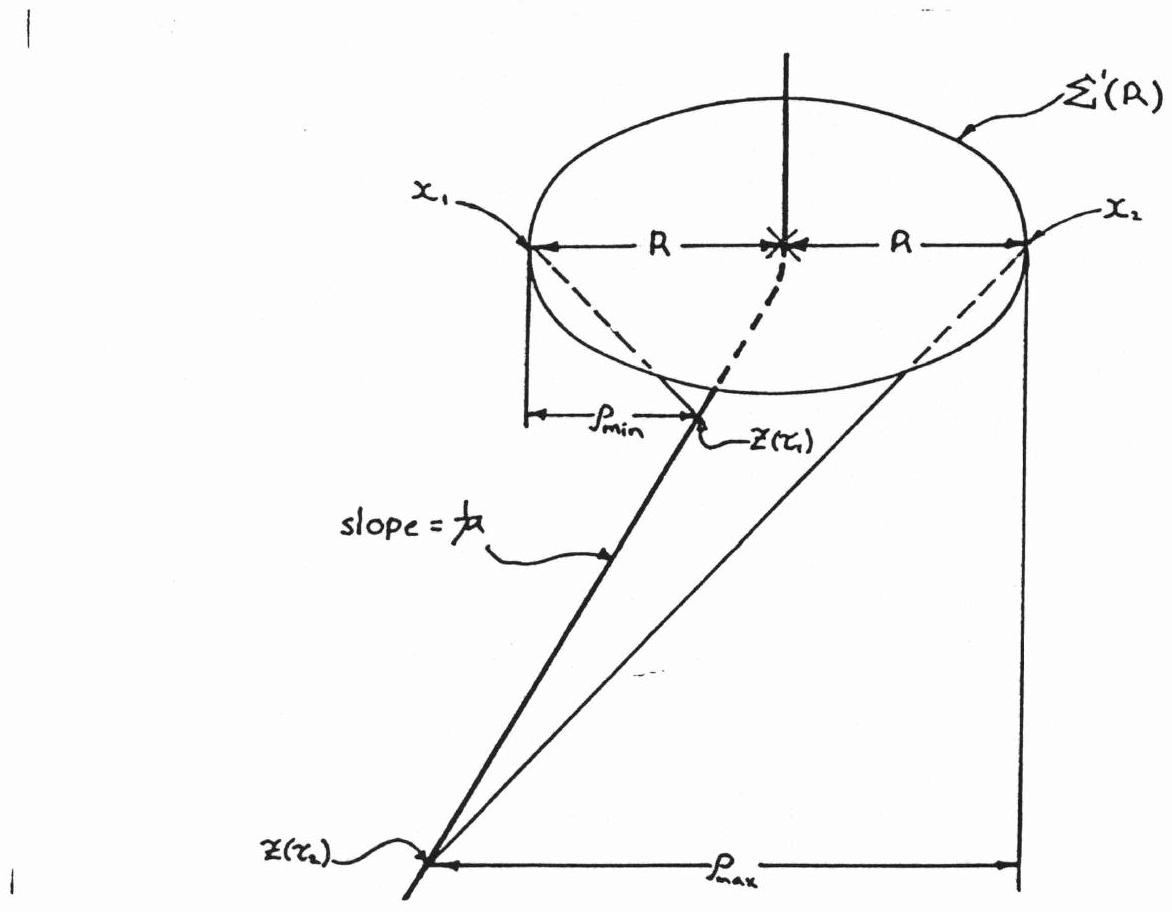}
  \caption{}
  \label{fig:4.5}
\end{figure}

The extreme values of $R/\rho$ will occur in a rest-frame of the charge in which the
charge has constant speed $\mu$ for its entire history, except for a brief moment
when it slowed to be at rest at the present time. This situation is pictured in
Figure~\ref{fig:4.5}.

{
\setlength{\abovedisplayskip}{4pt plus 2pt minus 2pt}%
\setlength{\belowdisplayskip}{4pt plus 2pt minus 2pt}%
After some algebra we find
\begin{equation}
  \left(\frac{1-\mu}{1+\mu}\right)^{\frac{1}{2}}
  < \frac{R}{\rho}
  < \left(\frac{1+\mu}{1-\mu}\right)^{\frac{1}{2}}
  \label{eq:4.54}
\end{equation}
Thus, since we have chosen $\mu < 1$, $R/\rho$ is bounded. Writing out
explicitly the integral we are computing and making the substitution
for $T^{\mu\nu}_s$ from above we arrive at
\begin{equation}
  \lim_{R\to\infty}\int_{\Sigma'(R)} T^{\mu\nu}_s\,n_\nu\,R^2(1+Ra_n)\,d\Omega
  = \lim_{R\to\infty}\int_{\Sigma'(R)} \frac{R^3}{\rho^3}a_n B^{\mu\nu}n_\nu\,d\Omega.
  \label{eq:4.55}
\end{equation}

This is as far as we can go with the integration, without knowing the specific
nature of the world-line. However, since $R/\rho$ is bounded and non-zero, in
general the right-hand side of \eqref{eq:4.55} will only be zero when $B^{\mu\nu}$
is zero in the limit $R \to \infty$. Referring back to \eqref{eq:4.21} we see that
$B^{\mu\nu}$ is proportional to the retarded acceleration. Thus if the acceleration
of the particle is zero in the remote past then $B^{\mu\nu}$ and hence the integral
over $\Sigma'(R)$ will vanish in the limit $R \to \infty$.

So, assuming
\begin{equation}
  a^\mu(-\infty) = 0
  \label{eq:4.56}
\end{equation}
then we have
\begin{equation}
  \frac{dP^\mu_s}{d\tau}
  = \lim_{\epsilon\to0}\int_{\Sigma'(\epsilon)} T^{\mu\nu}_s\,n_\nu\,\epsilon^2(1+\epsilon a_n)\,d\Omega
  \label{eq:4.57}
\end{equation}
\enlargethispage{0.5\baselineskip}
In order to evaluate this integral, we expand the integrand in powers of $\epsilon$.
The calculations involved are extremely long and complicated. Teitelboim quotes the
result from \cite{Dirac1938} who uses a number of short-cuts to simplify the algebra.
We have repeated the calculations using \textit{Mathematica}, the details of which
may be found in appendix \ref{app:A}. Substituting \eqref{eq:A.13} into \eqref{eq:4.57}
and applying the following identities:
\begin{align}
  \int_{\Sigma'(\epsilon)} n^\alpha\,d^2\sigma &= 0,
  \label{eq:4.58}\\
  \int_{\Sigma'(\epsilon)} n^\alpha n^\beta n^\gamma\,d^2\sigma &= 0,
  \label{eq:4.59}\\
  \int_{\Sigma'(\epsilon)} n^\alpha n^\beta\,d^2\sigma
  &= \frac{4\pi\epsilon^2}{3}\left(\eta^{\alpha\beta} + v^\alpha v^\beta\right),
  \label{eq:4.60}
\end{align}
}
we arrive at
\begin{equation}
  \frac{dP^\mu_s}{d\tau} = \lim_{\epsilon\to0}\left(\frac{e^2}{2\epsilon}a^\mu\right) - \frac{2}{3}e^2\dot{a}^\mu.
  \label{eq:4.61}
\end{equation}

The terms vanishing with $\epsilon$ have been dropped from \eqref{eq:4.61}. Notice that
this integral is in fact divergent, as we would expect since $P^\mu_s(\tau)$ contains
the four-momentum of the Coulomb field.

To find $P^\mu_s(\tau)$ we must integrate \eqref{eq:4.61} and hence we require an initial
condition. Equation \eqref{eq:4.56} implies that the field over all space-time at $\tau = -\infty$
is purely Coulombian, and so for the rest-frame we can write
\begin{equation}
  P^\mu_s(-\infty) = -\int T^{\mu\nu}_I\,v_\nu(-\infty)\,d^3x.
  \label{eq:4.62}
\end{equation}

Substituting the expressions we derived in Chapter~\ref{chap:3} for $r^\mu$, $v^\mu$, etc,
(Eqs.\ \eqref{eq:3.10}--\eqref{eq:3.14}) into \eqref{eq:4.20}, the integral reduces to
\begin{equation}
  P^\mu_s(-\infty)
  = \lim_{\epsilon\to0}\int_\epsilon^\infty R^2\,dR\int_0^{2\pi} d\phi\int_0^\pi \sin\theta d\theta\,
    \frac{e^2}{8\pi R^4}(1,\mbf{0})
  \label{eq:4.63}
\end{equation}
where $\mbf{n}$ is the unit vector $(\sin\theta\cos\phi,\sin\theta\sin\phi,\cos\theta)$.
Performing the integration, and recalling that we are in the frame where $v^\mu = (1,\mbf{0})$,
we have for an arbitrary frame
\begin{equation}
  P^\mu_s(-\infty) = \lim_{\epsilon\to0}\frac{e^2}{2\epsilon}v^\mu(-\infty)
  \label{eq:4.64}
\end{equation}
With this result and the fact that $a^\mu(-\infty)=0$, we can integrate \eqref{eq:4.61}
to obtain
\begin{equation}
  P^\mu_s(\tau) = \lim_{\epsilon\to0}\frac{e^2}{2\epsilon}v^\mu(\tau) - \frac{2}{3}e^2 a^\mu(\tau).
  \label{eq:4.65}
\end{equation}

Thus the bound four-momentum at a given instant has the remarkable
property that it depends only upon the four-velocity and
four-acceleration at the same time, even though the value of
$T^{\mu\nu}_s$ on $\sigma(\tau)$ is dependent upon the whole history
of the world-line prior to $\tau$. This should be compared with our
expression for $P^\mu_{II}$, which does depend on the whole history of
the charge. This is a strong justification for deeming $P^\mu_s$ to be
`bound' to the charge.

In deriving \eqref{eq:4.65} we had to invoke the boundary condition
\eqref{eq:4.56} and so this expression for the bound four-momentum can
only be considered valid in the case of a world-line with uniform
motion in the remote past. This is a curious boundary condition to
place on the charge, and one wonders if it is not simply an artefact
of our derivation process. In a paper by M. Sorg \cite{Sorg1978} it is
shown through an alternative calculation that \eqref{eq:4.65} is in
fact valid under the much weaker condition:
\begin{equation}
  \lim_{\tau'\to-\infty}\frac{v^\mu(\tau)}{(z^\nu(\tau)-z^\nu(\tau'))v_\nu(\tau)} = 0.
  \label{eq:4.66}
\end{equation}
This is a condition upon $v^\mu(-\infty)$, but it is much weaker than
the asymptotic condition \eqref{eq:4.56} of uniform motion in the
distant past. For instance, every world-line with bounded
four-velocity as in \eqref{eq:4.53} satisfies \eqref{eq:4.66}.

\subsection{Equation of Motion}
\label{sec:4.3.5}

{
\setlength{\abovedisplayskip}{4pt plus 2pt minus 2pt}%
\setlength{\belowdisplayskip}{4pt plus 2pt minus 2pt}%
Now that we know the bound and emitted four-momenta for a point charge
in arbitrary motion, we can calculate the equation of motion. Since
the particle cannot be separated from its bound electromagnetic
four-momentum (unlike the emitted electromagnetic four-momentum which
radiates away from the charge as soon as it is generated), we say that
the \textit{intrinsic} four-momentum of the charge is the sum of its
mechanical or bare\footnote{The expression `bare' is used because this
is the four-momentum that would be seen if we could `turn off' the
charge on our particle, i.e. observe the particle without its
associated electromagnetic field. In this case the observed mass would
be the `bare' mass.} four-momentum and its bound
four-momentum. Accordingly we write
\begin{equation}
P^\mu_{int} = P^\mu_s + P^\mu_{(bare)}.
\label{eq:4.67}
\end{equation}
If we assume the bare four-momentum to have the usual form for an
uncharged particle, then
\begin{equation}
P^\mu_{int} = \left(m_{(bare)} + \lim_{\epsilon\to 0}\frac{e^2}{2\epsilon}\right)v^\mu - \frac{2}{3}e^2 a^\mu.
\label{eq:4.68}
\end{equation}
The divergent term $e^2/2\epsilon$ arises unavoidably because of the
point-like nature of the charge.

To handle the divergence, we make the same identification as Dirac in
the previous section, namely
\begin{equation}
m = \left(m_{(bare)} + \lim_{\epsilon\to 0}\frac{e^2}{2\epsilon}\right).
\label{eq:4.69}
\end{equation}
The identification above reflects the empirical truth that the
rest-mass $m$ of a point charged particle (such as the electron) is
finite. The implication is: the electron has a negatively infinite
bare mass $m_{(bare)}$, so defined that, when it is added to the
infinite positive mass of the surrounding Coulomb field, the result is
the usual rest-mass of the electron. We are well aware that such a
description of the electron is extremely dubious, but in Teitelboim's
words: `No one has been able to make clear the inner physics of this
finiteness (of the electron's mass)\dots' In Chapter~\ref{chap:6} we
present an attempt of our own to avoid the mass-renormalization
problem.
}

So for the intrinsic four-momentum of our point charge we have
\begin{equation}
P^\mu_{int} = mv^\mu - \frac{2}{3}e^2 a^\mu.
\label{eq:4.70}
\end{equation}
Notice that in the rest-frame we get $P^0_{int} = m$ and so the
rest-energy of the particle is equal to its mass for a general motion.

To find the equation of motion for the particle in the absence of any
external force, we demand conservation of momentum for the system of
particle plus radiation; i.e.
\begin{equation*}
\frac{dP^\mu_{int}}{d\tau} = -\frac{dP^\mu_{II}}{d\tau}.
\end{equation*}
Substituting in for $P^\mu_s$ and $P^\mu_{II}$ we find
\begin{equation}
ma^\mu - \frac{2}{3}e^2\dot{a}^\mu = -\frac{2}{3}e^2a^2v^\mu.
\label{eq:4.71}
\end{equation}
Of course, when the particle is acted on by an external four-force
$F^\mu$, \eqref{eq:4.71} must become\footnote{If we make the
definition $P^\mu_{em} = P^\mu_s + P^\mu_{II}$ then \eqref{eq:4.72} is
simply the statement
\begin{equation*}
\frac{dP^\mu_{bare}}{d\tau} + \frac{dP^\mu_{em}}{d\tau} = F^\mu;
\end{equation*}
i.e. the rate of change of field momentum and mechanical momentum is
equal to the applied force.\label{foot:4}}
\begin{equation}
ma^\mu - \frac{2}{3}e^2\dot{a}^\mu = -\frac{2}{3}e^2a^2v^\mu + F^\mu.
\label{eq:4.72}
\end{equation}
This is the \textit{Lorentz-Dirac Equation}.\footnote{Our previous
expression derived via Dirac's technique had the four-force term as
$F^\mu = eF^{\mu\nu}_{in}v_\nu$, where $F^{\mu\nu}_{in}$ represented
electromagnetic waves incident on the charge. In the present
derivation we allow $F^\mu$ to be caused by other forces than
electromagnetism.}

\subsection{Critique of bound four-momentum}
\label{sec:4.3.6}

At this juncture it is worth re-examining Teitelboim's definition of
the bound electromagnetic four-momentum, $P^\mu_s$
\eqref{eq:4.37}. The bound four-momentum is derived from an integral
of the bound energy-momentum tensor over three-space in the rest-frame
of the charge at any instant. Its ``bound'' character is exemplified
by the fact that it depends on the surface of integration, in contrast
to the emitted four-momentum which was found to be
surface-independent. This observation was the motivation for splitting
the energy-momentum tensor into two parts, one which contained the
radiation field of the charge and the other containing the bound or
Coulomb field. Following through with this definition we found the
bound electromagnetic four-momentum to consist of a divergent
velocity-dependent part, and a finite acceleration-dependent term
(Eq.~\ref{eq:4.65}). The acceleration term is a rather curious addition to
what we normally regard as ``mechanical'' four-momentum.  It does not
correspond to an extra mass of the charge, but is more like a kind of
\emph{acceleration energy} (a term coined by Schott). Thus its
inclusion as part of the bound four-momentum seems a little
unjustified; we expect bound four-momentum to be derived from mass
that is effectively attached to the charge and is thereby proportional
to the four-velocity. The only real justification for including the
acceleration term as part of the bound four-momentum is the fact that
$P^\mu_s$ is surface dependent.

One is led to ponder, however, the possibility that the acceleration
term in $P^\mu_s$ might be invariant under changes in the surface, and
hence could itself be identified as a kind of emitted
four-momentum. It would be of a different character to $P^\mu_{II}$
because it would not be detectable as radiation, and in fact it is
unlikely that it would be derivable from a single part of the
energy-momentum tensor, as $P^\mu_{II}$ is. However we can choose
whatever criterion we like for defining radiated four-momentum, and
invariance under transformations in the surface of integration is
perhaps the most relativistically sensible definition.

Having said all this, it is in fact true that the acceleration term is
unaltered by a change in the surface, as is shown in
\cite{TeitelboimReviewUnknown}. For simplicity we consider what
happens when the surface of integration $\sigma(\tau)$ is
tilted\footnote{As the calculation of four-momentum is dependent only
on the local behaviour of the surface near the charge, and locally the
surface is a plane, the result above for a surface of constant normal
$u^\mu$ will be the same for an arbitrary surface whose normal at
$z(\tau)$ is $u^\mu$.}.  We denote by $u^\mu$ the normal to the
surface. Quoting \cite{TeitelboimReviewUnknown} we have for the
modified bound four-momentum
\begin{equation}
  P^{\prime\mu}_{self}(\tau) = e^2\left( \lim_{\epsilon\to 0}\left(\frac{2}{3\epsilon}\right)v^\mu
  + \frac{1}{6\epsilon}\left(\frac{u^\mu}{v^\nu u_\nu}\right) - \frac{2}{3}a^\mu \right),
  \label{eq:4.73}
\end{equation}
which clearly shows that the acceleration term is not affected by a
change in the surface. If we set $u^\mu = v^\mu$ then \eqref{eq:4.73}
reduces to \eqref{eq:4.65} as it should. Thus we claim that the true
bound electromagnetic four-momentum is the velocity-dependent,
divergent part of \eqref{eq:4.65}, whilst the acceleration term in
\eqref{eq:4.65} is to be included in the emitted four-momentum,
$P^\mu_{II}$. Such an identification does not alter the equation of
motion; it serves only to improve our understanding of the
characterisation of electromagnetic four-momentum into bound and
emitted parts.

\section{Conclusion}
\label{sec:4.4}

Through the use of only retarded fields we have managed to derive the
Lorentz-Dirac equation, thus removing some of the objections to the
original Dirac procedure. There still, however, remains the problem of
mass renormalization and how one is to interpret the supposedly
infinite bare mass of the charge. In Chapter~\ref{chap:6} we give an
alternative derivation of the equation where it is not necessary to
perform any kind of mass renormalisation. In the meantime, though, we
turn our attention to the possibility that a point-charge interacts
with its own field, and examine what implications this has for the
equation of motion.

\chapter{SELF-INTERACTION}
\label{chap:5}

\section{Background}
\label{sec:5.1}

In Dirac's expression for the Lorentz-Dirac equation~\eqref{eq:4.14}
the external electromagnetic field $F^{\mu\nu}_{in}$ appears
explicitly, whereas the self-field of the charge does not. It is
surprising that these fields play such asymmetrical roles in the
equation of motion for they are both defined at the point of the
charge (the self-field albeit having the value infinity there) and so
we would expect the charge to move under the influence of both
fields. Additionally, although we have used the balance of
four-momentum between particle and field to derive the equation of
motion, we have given no explanation of how a point charge ``knows''
to change its material momentum in accordance with changes in its
field momentum. It would be more intuitively acceptable if the
equation of motion could be derived by considering only the action of
fields on the particle.

In this chapter we show that the charge can indeed be thought of as
interacting with its own field, and that the Lorentz-Dirac equation is
nothing more than the usual Lorentz force law
(\textit{cf.}~\eqref{eq:3.28}) with the self-field of the charge
properly taken into account.

\setstretch{1}
\section{Justification of self-force in the equation of motion}
\setstretch{\bodystretch}
\label{sec:5.2}
We postulate that a point charge under the influence of an external
electromagnetic field $F^{\mu\nu}_{ext}$ moves according to the
\textit{modified Lorentz-Force equation}
{
\setlength{\abovedisplayskip}{4pt plus 2pt minus 2pt}%
\setlength{\belowdisplayskip}{4pt plus 2pt minus 2pt}%
\begin{equation}
  m a^\mu = e\left(F^{\mu\nu}_{ext} + F^{\mu\nu}_{ret}\right) v_\nu.
  \label{eq:5.1}
\end{equation}
$F^{\mu\nu}_{ret}$ is the retarded self-field evaluated at the
position of the charge.

Before we go into the derivation of the equation of motion we want to
demonstrate that the inclusion of the self-field is by no means an
\textit{ad hoc} assumption. In~\cite{Teitelboim1971} Teitelboim gives
an heuristic argument for the inclusion of the self-field, and
both~\cite{Barut1964} and~\cite{Rohrlich1965} use the self field to
derive the equation of motion through an action principle. However
nowhere in the literature does there appear a proof that four-momentum
balance and self-interaction should produce the same equation of
motion. We provide a proof of this here. In fact we will prove the
following relation:
\begin{equation}
  \frac{dP^\mu_{em}}{d\tau} = -eF^{\mu\nu}_{ret}v_\nu,
  \label{eq:5.2}
\end{equation}
where $P^\mu_{em}$ is the electromagnetic four-momentum ($=P^\mu_s +
P^\mu_{II}$ in \S\ref{sec:4.3.2}).
Substituting~\eqref{eq:5.2} into~\eqref{eq:5.1} we find
\begin{equation}
  m a^\mu + \frac{dP^\mu_{em}}{d\tau} = eF^{\mu\nu}_{ext}v_\nu
  \label{eq:5.3}
\end{equation}
which yields the equation of motion from four-momentum balance (see
the footnote on page 61).

\begin{figure}
  \vspace*{-1.8cm}
  \centering
  \includegraphics[width=0.8\textwidth]{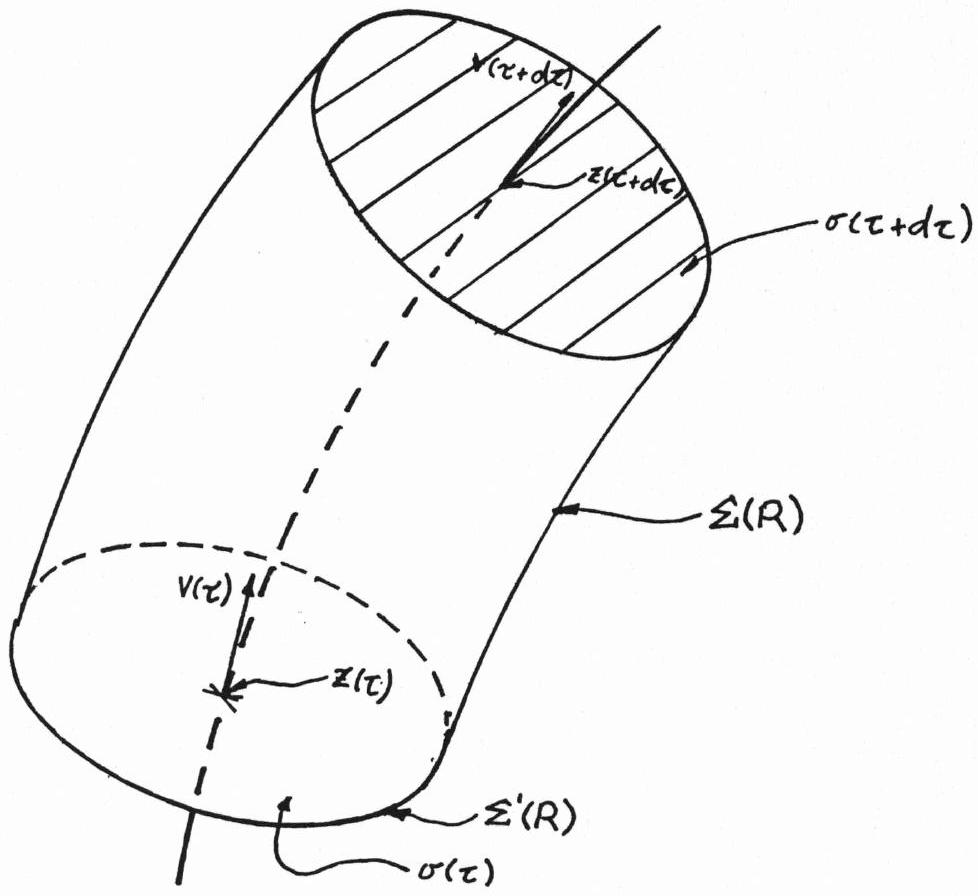}
  \caption{Calculation of $\dfrac{dP^\mu_{em}}{d\tau}$ in terms of the
    self-field.}
  \label{fig:5.1}
\end{figure}

The proof of~\eqref{eq:5.2} proceeds thus: We surround the world-line
of the charge by a tube of invariant radius $R$, which we denote by
$\Sigma(R)$ as in \S\ref{sec:4.2.2}. We then apply Gauss' theorem to
the region of space-time bounded by $\Sigma(R)$ and two surfaces
orthogonal to the world line at $z(\tau)$ and $z(\tau+d\tau)$
($\sigma(\tau)$ and $\sigma(\tau+d\tau)$, see
Figure~\ref{fig:5.1}). Recalling
definitions~\eqref{eq:4.33}\footnote{Note that we have specified the
arbitrary surface $S$ in~\eqref{eq:4.33} to be the orthogonal plane at
any world line point, as we are entitled to do since the definition is
independent of $S$.} and~\eqref{eq:4.37}, and taking the limit as $R
\to \infty$ we find
\begin{equation}
  P^\mu_{em}(\tau+d\tau) - P^\mu_{em}(\tau)
  + \lim_{R\to\infty}\int_{\Sigma(R)} T^{\mu\nu} n_\nu \, d^3\sigma
  = d\tau \int_{\sigma(\tau)} T^{\mu\nu}{}_{,\nu}\, d^3\sigma.
  \label{eq:5.4}
\end{equation}
We can rewrite the integral on the right-hand side of~\eqref{eq:5.4}
as follows
\begin{equation}
  \int_{\sigma(\tau)} T^{\mu\nu}{}_{,\nu}\, d^3\sigma
  = \int \delta\!\left[v^\alpha(\tau)\left(x_\alpha - z_\alpha(\tau)\right)\right] T^{\mu\nu}{}_{,\nu}\, d^4x,
  \label{eq:5.5}
\end{equation}
where the integration is now performed over the whole of space-time
(the delta function serves to restrict the space-time integration to
the plane $\sigma(\tau)$).
Substituting~\eqref{eq:5.5} into~\eqref{eq:5.4}, recalling from
\S\ref{sec:4.3.4} that the third term in the left-hand side
of~\eqref{eq:5.4} is zero under suitable conditions on the world line,
and applying~\eqref{eq:3.6} we find
\begin{equation}
  P^\mu_{em}(\tau+d\tau) - P^\mu_{em}(\tau)
  = -d\tau \int \delta\!\left[v^\alpha(\tau)\left(x_\alpha - z_\alpha(\tau)\right)\right] F^{\mu\nu}(x) j_\nu(x)\, d^4x.
  \label{eq:5.6}
\end{equation}
Using the form~\eqref{eq:2.10} for $j^\mu$ from Chapter~\ref{chap:2},
and noting that $d\tau$ is infinitesimal, we get
\begin{equation}
  \frac{dP^\mu_{em}(\tau)}{d\tau}
  = -e \int \delta\!\left[v^\alpha(\tau)\left(x_\alpha - z_\alpha(\tau)\right)\right] F^{\mu\nu}(x)
  \int_{-\infty}^{\infty} \delta^4\!\left(x - z(\tau')\right)\frac{dz_\nu}{d\tau'}\, d\tau'\, d^4x.
  \label{eq:5.7}
\end{equation}
Performing the $x$ integration yields
\begin{equation}
  \frac{dP^\mu_{em}(\tau)}{d\tau}
  = -e \int_{-\infty}^{\infty} \delta\!\left[v^\alpha(\tau)\left(z_\alpha(\tau') - z_\alpha(\tau)\right)\right]
  F^{\mu\nu}\!\left(z(\tau')\right)\frac{dz_\nu}{d\tau'}\, d\tau',
  \label{eq:5.8}
\end{equation}
which, on account of the identity
\begin{equation}
  \int_{-\infty}^{\infty} \delta\!\left[f(x)\right] g(x)\, dx
  = \left.\frac{g(a)}{|f'(a)|}\right|_{a=f^{-1}(0)},
  \label{eq:5.9}
\end{equation}
reduces to the expression we set out to prove
\begin{equation}
  \frac{dP^\mu_{em}(\tau)}{d\tau} = -eF^{\mu\nu}\!\left(z(\tau)\right)v_\nu(\tau).
  \label{eq:5.10}
\end{equation}
}

It must be stressed that this proof is in a sense only ``formal'', for
both the left-hand side and the right-hand side of~\eqref{eq:5.10} are
divergent quantities. A more rigorous proof awaits the advent of
mathematics capable of coping with the infinities inherent in this
subject.

\section{Self-field through retarded averaging}
\label{sec:5.3}

There are two major problems associated with defining the self-field
of a point charge on the world line. Firstly, the field is infinite at
the world line, and secondly the ``limit'' depends on the direction of
approach to the world line. The divergence problem we should be able
to get around by a Taylor expansion of the field near the world line,
and then absorption of any divergent terms through mass
renormalization as in the previous chapter. The problem with the limit
can also be circumvented by averaging the field over all possible
directions. This is Teitelboim's technique~\cite{Teitelboim1971},
which we describe below.

At a given instant we take the value of the field on the world line to
be the limit as $\epsilon$ tends to zero of the integral of
$F^{\mu\nu}_{ret}$ over the surface of a sphere of radius $\epsilon$
centered on the charge in its instantaneous rest frame, divided by
$4\pi\epsilon^2$. To find the value in any other frame we Lorentz
transform the rest frame result. The rest frame of the charge again
plays a privileged role as it is from this frame that the charge
``sees'' the universe.

\begin{figure}[!t]
  \vspace*{-2cm}
  \centering
  \includegraphics[width=0.95\textwidth]{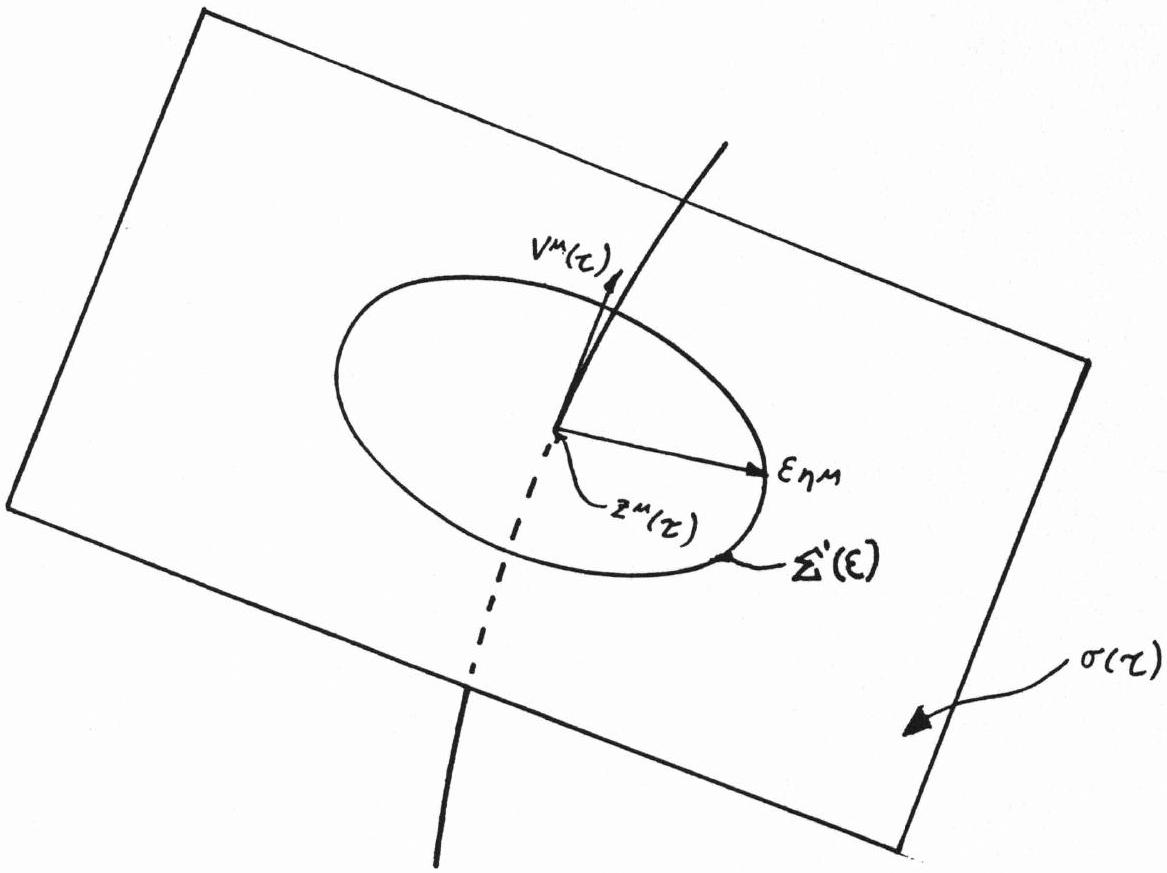}
  \caption{Calculation of retarded self-field. $\Sigma'(\epsilon)$ is
    the intersection of a tube of radius $\epsilon$ surrounding the
    world-line and the space-like surface orthogonal to the world-line
    at $z(\tau)$. The vector from $z(\tau)$ to a generic point on
    $\Sigma'(\epsilon)$ is written as $\epsilon n^\mu$, with $n^2=1$.}
  \label{fig:5.2}
\end{figure}

In geometrical terms this translates to an integration over
$\Sigma'(\epsilon)$ as defined in the last chapter, and as shown in
Figure~\ref{fig:5.2}. With the notation introduced in
Figure~\ref{fig:5.2} the value $F^{\mu\nu}_{ret}(z)$ of the retarded field at point $z$ on
the world line of the charge is
{
\setlength{\abovedisplayskip}{4pt plus 2pt minus 2pt}%
\setlength{\belowdisplayskip}{4pt plus 2pt minus 2pt}%
\begin{equation}
  F^{\mu\nu}_{ret}(z) = \lim_{\epsilon\to0}\frac{1}{4\pi\epsilon^2}\int_{\Sigma'(\epsilon)} F^{\mu\nu}_{ret}(z+\epsilon n)\, d^2\sigma.
  \label{eq:5.11}
\end{equation}
Dirac~\cite{Dirac1938} found an expansion for
$F^{\mu\nu}_{ret}(z+\epsilon u)$ in the appendix to his paper, which
Teitelboim borrows. However, Dirac uses a different metric to
Teitelboim and so the conversion between the two is not trivial. In
fact it appears that Teitelboim is in error for upon repeating the
calculation using Teitelboim's conventions and \textit{Mathematica},
we get a result which contains an extra term. A description of the
calculation is given in appendix~\ref{app:A}. The result of the
calculation (\textit{cf.}~\eqref{eq:A.12}) is
\begin{align}
  F^{\mu\nu}_{ret}(z+\epsilon n)
  &= e\left[
    \frac{1}{\epsilon^2}v^{[\mu}n^{\nu]}
    + \frac{1}{2\epsilon}\left(a^{[\mu}v^{\nu]} + a_n n^{[\mu}v^{\nu]}\right)\right. \notag\\
   &+ \left.\frac{3}{4}a_n v^{[\mu}a^{\nu]}
    + \frac{1}{8}a^2 v^{[\mu}n^{\nu]}
    + \frac{1}{2}n^{[\mu}\dot a^{\nu]}\right. \notag \\
   &+ \left.\frac{2}{3}v^{[\mu}\dot a^{\nu]}
    + \frac{3}{8}a_n^2 v^{[\mu}n^{\nu]}
    + O(\epsilon)\right].\label{eq:5.12} 
\end{align}
The extra term here is $\frac{3}{8}a_n^2 v^{[\mu}n^{\nu]}$.

Substituting~\eqref{eq:5.12} into~\eqref{eq:5.11} and applying the
identities~\eqref{eq:4.58}--\eqref{eq:4.60} we see that the extra term
in~\eqref{eq:5.12} integrates to zero, and so we obtain the same
result as Teitelboim:
\begin{equation}
  F^{\mu\nu}_{ret}(z)
  = e\left[
    -\left(\lim_{\epsilon\to0}\frac{2}{3\epsilon}\right)v^{[\mu}a^{\nu]}
    - \frac{2}{3}\dot a^{[\mu}v^{\nu]}
  \right].
  \label{eq:5.13}
\end{equation}
Contracting~\eqref{eq:5.13} with $v^\mu$ yields the Lorentz force:
\begin{equation}
  eF^{\mu\nu}_{ret}v_\nu
  = -\left(\lim_{\epsilon\to0}\frac{2e^2}{3\epsilon}\right)a^\mu
    + \frac{2}{3}e^2\left(\dot a^\mu - a^2 v^\mu\right).
  \label{eq:5.14}
\end{equation}
}
Substituting~\eqref{eq:5.14} into~\eqref{eq:5.1}, and
mass-renormalizing\footnote{It is of no consequence that the divergent
term to be absorbed in this case is four thirds of the divergent term
in the last chapter, as we are taking the limit $\epsilon \to 0$ and
there is no reason to suppose that the $\epsilon$'s can be compared.}
the divergent term we arrive again at the Lorentz-Dirac equation.

The only arbitrary feature of the preceding derivation is the
definition of $F^{\mu\nu}_{ret}$ on the world line. There is no
convincing reason why the rest frame of the charge should be so
important in this definition, and so it is interesting to contemplate
the possibility of alternative definitions, for example we could tilt
the plane of integration in~\eqref{eq:5.11}. Teitelboim addresses this
question briefly in his paper, claiming but not proving that
definition~\eqref{eq:5.11} is unique if we require the field on the
world-line to be ``maximally matched'' with the field off the world
line. He does not define what he means by ``maximally matched'' and so
we are left wondering as to the truth of his statement.

In the next section we consider an alternative method for defining the
field on the world line, due to Barut \cite{Barut1974} (and discovered
independently by the author) which leads to a very quick derivation of
the Lorentz-Dirac equation.

\section{Self-field via analytic continuation}\label{sec:5.4}

We define the field on the world line of a point charge by
\textit{analytic continuation} of the retarded field of the charge. We
consider the retarded field to be a function of \textit{two}
variables, $F(x;z(\tau))$\footnote{Usually we require
$(x-z(\tau))^2=0$ to hold, thus fixing $z(\tau)$ if we are given $x$.}
and continue analytically Eq.~\eqref{eq:2.22} to the unphysical point
$x=z(\tau)$, $z=z(\tau-\epsilon)$. The field at $x=z(\tau)$ is then
taken to be the limit as $\epsilon \to 0$.

The calculation is performed in Appendix~\ref{app:B}, and after
substitution of the field $F^{\mu\nu}_{ret}$ as calculated there
into~\eqref{eq:5.1} we arrive at
\begin{equation}
  \left(m+\lim_{\epsilon\to0}\left(\frac{e^2}{2\epsilon}\right)\right)a^\mu
  = \frac{2e^2}{3}\left(\dot a^\mu - a^2 v^\mu\right) + eF^{\mu\nu}_{ext}v_\nu,
  \label{eq:5.15}
\end{equation}
which after mass-renormalization becomes the Lorentz-Dirac equation.

This technique is particularly useful as it does not require the
calculation of $F^{\mu\nu}_{ret}$ at arbitrary space-time points. In
fact Barut~\cite{BarutVillarroel1975a,BarutVillarroel1975b} has used
this method to derive the Lorentz-Dirac equation for a charge in
curved space-time, and it considerably simplifies the usually
laborious calculations involved.

\chapter{A DISCUSSION OF RENORMALIZATION}
\label{chap:6}

\section{The problems of mass-renormalization}
\label{sec:6.1}

In all of the equation-of-motion derivations we have presented so far
an \emph{ad hoc} renormalization of the mass was necessary to achieve
a physically reasonable equation. Unfortunately, both physically and
mathematically, mass renormalization is an unacceptable procedure. We
present below a few of the inconsistencies caused by
mass-renormalization.

\subsection{General Relativity}
\label{sec:6.1.1}

We would expect the idea that an electron has a negatively infinite
bare mass at its center to cause problems with the electron's
gravitational metric. This is indeed the case as we shall see below.

The Reissner-Nordstr\"om\footnote{For a discussion of some of the
curious properties of this metric,
see~\cite[p.~156--161]{HawkingEllis73}.}  metric represents the
space-time in a region surrounding a spherically symmetric charged
body. There exist coordinates in which the metric has the form
{
\setlength{\abovedisplayskip}{4pt plus 2pt minus 2pt}%
\setlength{\belowdisplayskip}{4pt plus 2pt minus 2pt}%
\begin{equation}\label{eq:6.1}
ds^{2}
= -\left(1-\frac{2m}{r}+\frac{e^{2}}{r^{2}}\right)dt^{2}
+ \left(1-\frac{2m}{r}+\frac{e^{2}}{r^{2}}\right)^{-1}dr^{2}
+ r^{2}d\Omega^{2},
\end{equation}
where $m$ is the gravitational mass and $e$ is the charge of the body.

We contrast this solution to the Schwarzschild metric which is the metric
for an uncharged spherically-symmetric body, and which takes the form
\begin{equation}\label{eq:6.2}
ds^{2}
= -\left(1-\frac{2m}{r}\right)dt^{2}
+ \left(1-\frac{2m}{r}\right)^{-1}dr^{2}
+ r^{2}d\Omega^{2}.
\end{equation}
}

From \eqref{eq:6.1} and \eqref{eq:6.2} it is evident that the Coulomb
field modifies the gravitational potential of an uncharged body by a
factor of $e^{2}/r^{2}$.  At large distances (large $r$) where the
$-2m/r$ term is dominant the metric behaves as that of an uncharged
body, however near to the charge the $e^{2}/r^{2}$ term becomes
dominant, and in fact causes the gravitational field of the charge to
be repulsive at close range. Now, the $m$ appearing in \eqref{eq:6.1}
is the bare-mass of the charge and it was for inertial, not
gravitational, reasons that we had to postulate its value as
$-\infty$. However clearly even a finite negative mass would cause the
gravitational field of the charge to be repulsive at all
distances. Thus mass-renormalization would appear to be
phenomenologically inconsistent with general relativity, although a
more complete discussion using the gravitational energy-momentum
pseudotensor (see~\cite[p.~316-323]{LandauLifshitz80}) would be
necessary before any rigorous conclusions could be drawn.

\subsection{Mathematical problems}
\label{sec:6.1.2}

Since we use infinity as an ordinary real number when
mass-renormalizing, it is worth investigating what consequences this
has for the mathematics of the real line. We can consistently adjoin
$\pm\infty$ to the real line if we include the following rules
\begin{align*}
0\times\pm\infty &= 0\\
a\times\pm\infty &= \pm\infty \qquad (a>0)\\
a\times\pm\infty &= \mp\infty \qquad (a<0)\\
a+\infty &= \infty \qquad (a>-\infty).
\end{align*}
An important exclusion from these rules is a definition for
$\infty-\infty$. This is necessary for consistency; for example
$\lim_{n\to\infty}n=\infty$ and $\lim_{n\to\infty}n^{2}=\infty$,
however $\lim_{n\to\infty}(n^{2}-n)=\infty$ whereas
$\lim_{n\to\infty}(n-n)=0$ and so we cannot consistently define
$\infty-\infty$. The only way $\infty-\infty$ could be defined is if
we restrict its application to a single limiting procedure.

Now, mass-renormalization can be viewed as a subtraction of infinity
from infinity; the infinite bare mass of the point charge is
subtracted from its infinite Coulomb mass.  Therefore we could define
$\infty-\infty$ to be appropriate for the electron, say. i.e take
$\infty-\infty = 9.1\times 10^{-31}kg$. But what of other point-like
particles such as the muon that have the same charge as the electron
but a different mass? We would have to \emph{redefine} $\infty-\infty$
to be the mass of the muon. Thus because there exists more than one
point charged particle in nature, we cannot find an acceptable
mathematical procedure for mass-renormalization. It is probably better
to view mass-renormalization as merely an omission of divergent terms,
rather than the subtraction of two infinite quantities.

\section{Avoiding Mass-Renormalization}
\label{sec:6.2}

We now present a scheme for deriving the equation of motion of a point
charge that does away with the need for mass-renormalization.

We begin by turning our attention to the self-interaction of the
charge.  When self-interaction was introduced in the previous chapter,
it was done by modifying the Lorentz-Force equation \eqref{eq:3.28} to
include the self-force \eqref{eq:5.1}.  This approach is not quite
consistent with the four-momentum balance method of Chapter~\ref{chap:4},
for there we derived the equation of motion by equating the externally
applied force to the total rate of change of field four-momentum plus
particle four-momentum (see the footnote on page 61). It is only by
virtue of relation \eqref{eq:5.10} that the two approaches give the
same answer. An alternative approach is to form a synthesis of the two
methods; namely we claim that the total rate of change of field
four-momentum plus particle four-momentum is equal to the total
applied force, which includes the self-interaction of the charge. This
translates into the following equation
{
\setlength{\abovedisplayskip}{4pt plus 2pt minus 2pt}%
\setlength{\belowdisplayskip}{4pt plus 2pt minus 2pt}%
\begin{equation}\label{eq:6.3}
\frac{dP^\mu_{\mathrm{bare}}}{d\tau}+\frac{dP^\mu_{\mathrm{em}}}{d\tau}
= F^\mu_{\mathrm{ext}} + eF^{\mu\nu}_{\mathrm{self}}v_\nu.
\end{equation}

Previously (Chapter~\ref{chap:5}) we have taken the self-field
$F^{\mu\nu}_{\mathrm{self}}$ of the charge to be
$F^{\mu\nu}_{\mathrm{ret}}$. However clearly in the present
circumstance such a definition would not work as by \eqref{eq:5.10},
\eqref{eq:6.3} would become
\begin{equation}\label{eq:6.4}
\frac{dP^\mu_{\mathrm{bare}}}{d\tau}+2\frac{dP^\mu_{\mathrm{em}}}{d\tau}
= F^\mu_{\mathrm{ext}}
\end{equation}
and, among other things, would yield twice the correct radiation
rate. To rectify this problem we make the following definition for the
self-field
\begin{equation}\label{eq:6.5}
F^{\mu\nu}_{\mathrm{self}} = -\frac{1}{2}\left(F^{\mu\nu}_{\mathrm{ret}}+F^{\mu\nu}_{\mathrm{adv}}\right).
\end{equation}
}

This is a definition for the field on the world-line only. We have
already calculated $F^{\mu\nu}_{\mathrm{ret}}$ and
$F^{\mu\nu}_{\mathrm{adv}}$ on the world line in Appendix B, and so
substituting those expressions (Eqs.\ \eqref{eq:B.2} and
\eqref{eq:B.3}) into \eqref{eq:6.5} we arrive at
\begin{equation}\label{eq:6.6}
eF^{\mu\nu}_{\mathrm{self}}v_\nu = \lim_{\epsilon\to 0}\left(\frac{e^{2}}{2\epsilon}\right)a^\mu.
\end{equation}

Using \eqref{eq:6.5}, \eqref{eq:4.45} and \eqref{eq:4.61} (recalling
that $P^\mu_{\mathrm{em}}=P^\mu_{II}+P^\mu_s$), \eqref{eq:6.3} becomes
\begin{equation}\label{eq:6.7}
ma^\mu = F^\mu + e^{2}\frac{2}{3}\left(\dot{a}^{\mu}-a^{2}v^{\mu}\right)
\end{equation}
which is once again the Lorentz-Dirac equation. In this derivation we
have not needed to use mass-renormalization, and so $m$ is simply the
bare mass of the charge.

Some justification for the definition \eqref{eq:6.5} of the field on
the world line is warranted in view of its radical departure from the
definition of the field off the world line,
$(F^{\mu\nu}_{\mathrm{ret}})$. Firstly, we have introduced the
advanced field to describe the self-interaction of the particle, a
move of which we earlier we questioned the validity. However here we
have restricted its region of influence to the world line and so it
cannot give rise to any causality difficulties. In fact on the world
line of the charge the retarded and advanced fields have equal status,
as they both neither propagate forwards nor backwards in time there,
and so we would expect them to appear symmetrically in the self-field
of the particle. Secondly, the minus sign in the right-hand side of
\eqref{eq:6.5} would appear to be rather \emph{ad hoc}. In a sense it
is, for the only real justification we can give for its existence is
that it makes mass-renormalization unnecessary, and hence explains the
empirically finite mass of the electron.

Possibly we can view the interaction of a charge with itself, through
its own field as a process ``dual'' to the interaction of a charge
with an external field (in the former case a photon is emitted,
whereas in the latter case a photon is absorbed). In this way we might
expect a minus sign to appear.

\section{The Action Principle.}
\label{sec:6.3}

Most laws of physics can be expressed as an action principle, and we
would expect the equation of motion of a point charge to be no
different. Thus there should exist a Lagrangian from which the
equation of motion of a point charge is derivable.

The Lagrangian for a free (neutral) particle with world-line
$z(\lambda)$ is
{
\setlength{\abovedisplayskip}{4pt plus 2pt minus 2pt}%
\setlength{\belowdisplayskip}{4pt plus 2pt minus 2pt}%
\begin{equation}
I_{\mathrm{free}}=\int L_{\mathrm{free}}\,d\lambda=-m\int \sqrt{-\frac{dz^\mu}{d\lambda}\frac{dz_\mu}{d\lambda}}\,d\lambda
\label{eq:6.8}
\end{equation}
}
for then Lagrange's equations:
\begin{equation}
\frac{\partial L}{\partial z^\mu}=\frac{d}{d\lambda}
\left(\frac{\partial L}{\partial \dot{z}^\mu}\right)
\label{eq:6.9}
\end{equation}
(where $\dot{z}\equiv dz/d\lambda$), yield
\begin{equation}\label{eq:6.10}
m\ddot{z}^{\mu}=0.
\end{equation}
Here $I_{\mathrm{free}}$ is the length of the particle's world line,
and so the condition \eqref{eq:6.10} is simply telling us that the
shortest distance between two points is a straight line.

If we want to find the equation of motion for a charged particle we
must find a Lagrangian which takes into account the interaction
through its self field and any external fields.  Before embarking on a
search for such a Lagrangian we present the usual electromagnetic
Lagrangian and show how it yields Maxwell's equations. From this we
will get a hint for how to proceed with a charged particle. The
Lagrangian density for electromagnetism is
{
\setlength{\abovedisplayskip}{4pt plus 2pt minus 2pt}%
\setlength{\belowdisplayskip}{4pt plus 2pt minus 2pt}%
\begin{equation}\label{eq:6.11}
\mathcal{L}(x)=-\frac{1}{16\pi}F^{\mu\nu}(x)F_{\mu\nu}(x)+A^\mu(x)j_\mu(x)
\end{equation}
where, as usual, $F^{\mu\nu}(x)=A^{\nu,\mu}(x)-A^{\mu,\nu}(x)$. The
total action is found by integrating \eqref{eq:6.11} over the whole of
space-time
\begin{equation}\label{eq:6.12}
I_{\mathrm{field}}=\int \mathcal{L}(x)\,d^{4}x.
\end{equation}

The field equations corresponding to extremals of the action $I$ are
found by applying Lagrange's equations
\begin{equation}\label{eq:6.13}
\frac{\partial \mathcal{L}}{\partial A^\mu}=
\left(\frac{\partial \mathcal{L}}{\partial A^{\mu,\nu}}\right)_{,\nu}.
\end{equation}

Evaluating \eqref{eq:6.13} with \eqref{eq:6.11} results in Maxwell's
second equation \eqref{eq:2.2}, thus vindicating our choice of
Lagrangian.

Once Maxwell's equations have been derived from the action principle
we usually then go on to solve them for a particular current
distribution (e.g point charge). In the case of a point charge the
solution is the retarded and advanced fields of Chapter~\ref{chap:2}. Here we are
interested in quite the reverse problem: Given the field of a point
charge, what is its equation of motion?

The interaction term $A^\mu j_\mu$ in \eqref{eq:6.11} suggests how we
might modify \eqref{eq:6.8} for a charged particle. We simply write
\begin{equation}\label{eq:6.14}
I=I_{\mathrm{free}}+I_{\mathrm{int}} =\int
-m\sqrt{-\frac{dz^\mu}{d\lambda}\frac{dz_\mu}{d\lambda}}\,d\lambda +\int
\left(A^\mu(x)+A^\mu_{\mathrm{ext}}(x)\right)j_\mu(x)\,d^{4}x.
\end{equation}
The charge interacts with both external fields $A^\mu_{\mathrm{ext}}$
and its own field $A^\mu$. At this stage we do not specify the
particular combination of advanced and retarded potentials that make
up $A^\mu$; the necessary choice will become clear later on.

Substituting in the expression for a point charge current distribution
from Chapter~\ref{chap:2} \eqref{eq:2.10}, the interaction term in
\eqref{eq:6.14} becomes
\begin{equation}\label{eq:6.15}
I_{\mathrm{int}}
= e\int \left[A^\mu(x)+A^\mu_{\mathrm{ext}}(x)\right]\int\delta^{4}\!\left(x-z(\lambda)\right)
\dot z_\mu(\lambda)\,d\lambda\,d^{4}x.
\end{equation}
Performing the $x$ integration we arrive at
\begin{equation}\label{eq:6.16}
I_{\mathrm{int}}=e\int \left[A^\mu(z(\lambda))+A^\mu_{\mathrm{ext}}(z(\lambda))\right]\dot z_\mu(\lambda)\,d\lambda
\end{equation}
}
and so the total action for the charge is
\begin{equation}\label{eq:6.17}
I=\int -m\sqrt{-\frac{dz^\mu}{d\lambda}\frac{dz_\mu}{d\lambda}}
+e\left[A^\mu(z(\lambda))+A^\mu_{\mathrm{ext}}(z(\lambda))\right]\dot z_\mu(\lambda)\,d\lambda.
\end{equation}
  
Notice that if we take $A^\mu$ to be the retarded field of the
particle then the self-interaction term in \eqref{eq:6.17} will be
infinite. As the field lagrangian also contains such an interaction
term (recall \eqref{eq:6.11} and \eqref{eq:6.12}) this divergence
demonstrates rather a curious fact: If we start with the action
principle \eqref{eq:6.11}, solve for the potentials $A^\mu$ due to a
point charge, and then substitute the resulting expression for, say,
the retarded potential back into the total action, we find that the
total action is divergent. This problem is caused by the introduction
of time asymmetry into the theory, through the choice of
$F_{\mathrm{ret}}$ as the solution to Maxwell's equations. If instead
we choose the unique time symmetric combination of $F_{\mathrm{ret}}$
and $F_{\mathrm{adv}}$: $1/2(F_{\mathrm{ret}}-F_{\mathrm{adv}})$, then
the integral \eqref{eq:6.16} will be convergent and so will the total
action of the field\footnote{See in this context Rohrlich, Chapter~9.}
\eqref{eq:6.12}.

{
\setlength{\abovedisplayskip}{4pt plus 2pt minus 2pt}%
\setlength{\belowdisplayskip}{4pt plus 2pt minus 2pt}%
To find the equation of motion for a charged particle we apply
Lagrange's equations \eqref{eq:6.9} to the total action
\eqref{eq:6.17} and arrive at:
\begin{equation}\label{eq:6.18}
ma^\mu(\tau)=e\left[F^{\mu\nu}_{\mathrm{ext}}(z(\tau))+F^{\mu\nu}(z(\tau))\right]v_\nu(\tau).
\end{equation}

We still have not specified what $F^{\mu\nu}(z(\tau))$ is. This can be
determined by looking back at the preliminary form of the equation of
motion \eqref{eq:6.3} and substituting $ma^\mu$ for
$dP^\mu_{\mathrm{bare}}/d\tau$, our expression \eqref{eq:6.5} for
$F^{\mu\nu}_{\mathrm{self}}$ and recalling identity
\eqref{eq:5.2}. The resulting equation is
\begin{equation}\label{eq:6.19}
ma^\mu=\frac{e}{2}\left(F^{\mu\nu}_{\mathrm{ret}}-F^{\mu\nu}_{\mathrm{adv}}\right)v_\nu+F^\mu_{\mathrm{ext}}.
\end{equation}
}

Comparing this with \eqref{eq:6.18} we see that the correct definition
for $F^{\mu\nu}$ is
\begin{equation}\label{eq:6.20}
F^{\mu\nu}=\frac{1}{2}\left(F^{\mu\nu}_{\mathrm{ret}}-F^{\mu\nu}_{\mathrm{adv}}\right),
\end{equation}
which again leads to the Lorentz-Dirac equation without mass renormalization. The marked
difference between this definition for the self-field and definition \eqref{eq:6.5} is
due to the differing viewpoints lying behind their respective derivations. Eq.\
\eqref{eq:6.5} is derived by taking into account the momentum of the field of the charge,
whereas \eqref{eq:6.20} results simply from a modification of the action for a neutral
charge. In fact the latter approach is similar to that taken in perturbative Quantum
Electrodynamics where a point charge is first considered as a neutral particle, and then
the radiative corrections from its self field are added in. Modifying the action as we
have done above essentially amounts to treating the self field of the particle in the
same manner as the external field.

With definition \eqref{eq:6.20}, the action for a point charge
\eqref{eq:6.17} becomes
\begin{equation}\label{eq:6.21}
I=\int -m\sqrt{-\frac{dz^\mu}{d\lambda}\frac{dz_\mu}{d\lambda}}
+e\left[\frac{1}{2}\left[A^\mu_{\mathrm{ret}}(z(\lambda))-A^\mu_{\mathrm{adv}}(z(\lambda))\right]
+A^\mu_{\mathrm{ext}}(z(\lambda))\right]\dot z_\mu(\lambda)\,d\lambda.
\end{equation}

\section{Conclusion.}
\label{sec:6.4}

It is evident from the preceding discussion that the equation of
motion is derivable without mass-renormalization, but, to do so we
must introduce advanced fields on the world line of the
charge. Whether it is valid to propose that the only fields existing
off the world line are the retarded fields, whilst allowing advanced
fields on the world line of the charge, is a tricky question. Such an
approach causes the field to be highly discontinuous at the world
line, although it does explain the finite mass of the electron.  It is
worth pointing out that even with a definition like \eqref{eq:6.20}
for the self-field, the action \eqref{eq:6.12} will diverge for a
point charge if we take the field off the world line to be purely
$F_{\mathrm{ret}}$.

\chapter{CONCLUSION}
\label{chap:conclusion}
The historical dilemma faced by the founders of classical electrodynamics
was to choose between a structureless or structured electron. The idea of a
structured particle eventually had to be rejected on the grounds that it is
unstable because of its own coulombic repulsion, and also that its shape is
not relativistically invariant. On the other hand, we have seen that particles
without structure (point particles) have divergent self-energy which in turn
poses great mathematical and physical difficulties through the renormalization
process. In Chapter~\ref{chap:6} we addressed these problems and found that the
only way to avoid renormalization and hence explain the finite mass of the
electron was to include advanced fields when describing the self-interaction
of the particle. However, in order that causality not be violated we must
still retain only retarded fields off the world line, and so we saddle ourselves
with a field that is highly discontinuous at the world-line. Hopefully a more
thorough investigation would demonstrate whether or not such an approach
to classical electrodynamics could be formulated in a self-consistent way.

\appendix

\chapter{CALCULATION OF FIELD NEAR WORLD-LINE}
\label{app:A}
In this appendix we show how to calculate the field and the energy-momentum
tensor of a point charge near the world line. The bulk of the calculation is
performed using a \textit{Mathematica} program, however before we give a listing
of that program we first explain the problem.

\section{Preliminaries}
We consider a particle world line $z=z(\tau)$ with four-velocity $v^\mu$ and we
calculate its retarded field at space-time point $x^\mu$ where
\begin{equation}\label{eq:A.1}
x^\mu = z^\mu(\tau) + \epsilon n^\mu,
\end{equation}
with
\begin{align}
v^\mu(\tau)n_\mu &= 0, \label{eq:A.2}\\
n^\mu n_\mu &= 1.\label{eq:A.3}
\end{align}
The retarded field is given by~\eqref{eq:2.22}, which we reproduce
here for convenience:
{
\setlength{\abovedisplayskip}{3pt plus 2pt minus 2pt}%
\setlength{\belowdisplayskip}{3pt plus 2pt minus 2pt}%
\begin{equation}\label{eq:A.4}
F^{\mu\nu}_{\mathrm{ret}}
=
\frac{e}{\rho^3}\left(1+r_\sigma a^\sigma\right)v^{[\mu}r^{\nu]}
+
\frac{e}{\rho^2}a^{[\mu}r^{\nu]}.
\end{equation}
The same notation is used as in Chapter~\ref{chap:2}, namely
\begin{align}
r^\mu &= x^\mu - z^\mu(\tau_r), \label{eq:A.5}\\
\rho &= -\left(x^\mu - z^\mu(\tau_r)\right)v_\mu(\tau_r).\label{eq:A.6}
\end{align}
As usual, $\tau_r$ is the retarded world line point corresponding to $x^\mu$.
In this case we take
\begin{equation}\label{eq:A.7}
\tau_r = \tau - \delta.
\end{equation}
Using (\ref{eq:A.7}), all quantities dependent on the retarded proper time
$\tau_r$ may be expanded as power series in $\delta$. In this way we can find an
expression for $F^{\mu\nu}_{\mathrm{ret}}$ in terms of $\delta$ using
\textit{Mathematica}. The one thing \textit{Mathematica} can't do is find the
relationship between $\delta$ and $\epsilon$, which we need if we are to express
our result in terms of $\epsilon$.

To calculate the relationship between $\delta$ and $\epsilon$ we expand
$z^\mu(\tau_r)$ in a power series in $\delta$:
\begin{equation}\label{eq:A.8}
z^\mu(\tau_r)=z^\mu(\tau-\delta)=z^\mu-\delta v^\mu+\frac{\delta^2}{2}a^\mu
-\frac{\delta^3}{6}\dot a^\mu+O(\delta^4),
\end{equation}
where all terms in the right-hand side above are calculated at proper time
$\tau$. Now we use the fact that $r^\mu$ is a null vector,
\begin{equation}\label{eq:A.9}
\left(x^\mu-z^\mu(\tau_r)\right)\left(x_\mu-z_\mu(\tau_r)\right)=0,
\end{equation}
which combined with (\ref{eq:A.1}) and the
identities \eqref{eq:2.21}, \eqref{eq:2.20} yields
\begin{equation}
\epsilon^2-\delta^2(1+\epsilon a_n)+\frac{\epsilon\delta^3}{3}\dot a_n-\frac{\delta^4}{12}a^2=O(\epsilon^5).
\label{eq:A.10}
\end{equation}
}
Thus to first order $\delta=\epsilon$ and so without spoiling the accuracy of
(\ref{eq:A.10}) we can replace $\delta$ by $\epsilon$ in terms of order $\ge 3$.
Hence:
\begin{equation*}
\epsilon^2-\delta^2(1+\epsilon a_n)+\frac{\epsilon^4}{3}\dot a_n-\frac{\epsilon^4}{12}a^2=O(\epsilon^5).
\end{equation*}
Inverting and taking the square root results in the final expression
\begin{equation}
\delta=\epsilon\left(1+\frac{\epsilon^2}{6}\dot a_n-\frac{\epsilon^2}{24}a^2\right)
\left(1+\epsilon a_n\right)^{-\frac12}+O(\epsilon^4).
\label{eq:A.11}
\end{equation}
This expression appears in the \textit{Mathematica} program below.

In addition to calculating $F^{\mu\nu}$, the program also calculates
the energy-momentum tensor $T^{\mu\nu}$ and performs the contraction
in the integrand of~\eqref{eq:4.57}. The results of these calculations
are as follows:
\begin{align}
F^{\mu\nu}_{\mathrm{ret}}(z+\epsilon n)
&=
e\left[
\frac{1}{\epsilon^2}v^{[\mu}n^{\nu]}
+\frac{1}{2\epsilon}\left(a^{[\mu}v^{\nu]}+a_n n^{[\mu}v^{\nu]}\right)
+\frac{3}{4}a_n v^{[\mu}a^{\nu]}\right. \notag\\
&+\left.\frac{1}{8}a^2 v^{[\mu}n^{\nu]}
+\frac{1}{2}n^{[\mu}\dot a^{\nu]}
+\frac{2}{3}v^{[\mu}\dot a^{\nu]}\right.\notag\\
&+\left.\frac{3}{8}a_n^2 v^{[\mu}n^{\nu]}
+O(\epsilon)
\right].
\label{eq:A.12}
\end{align}
\begin{align}
T^{\mu\nu}_s(z+\epsilon n)n_\nu\,\epsilon^2(1+\epsilon a_n)
&=
\frac{e^2}{\pi}\left[
-\frac{1}{8\epsilon^2}n^\mu
+\frac{1}{8\epsilon}a^\mu
-\frac{3}{16}a_n a^\mu
-\frac{1}{6}\dot a^\mu\right. \notag\\
&\left.+\left(\frac{a_n^2}{4}-\frac{a^2}{8}\right)n^\mu
+\left(\frac{a_n^2}{4}-\frac{a^2}{12}\right)v^\mu
+O(\epsilon)
\right].\label{eq:A.13}
\end{align}

Note that in the program $y=\delta$, $k=r$ and $kar\equiv \rho a_r$.

\section{The Program}

\begingroup
\begin{verbatim}
(* This program calculates the electromagnetic field due to
   a point charge in arbitrary motion, close to the world line
   of the charge. The field is calculated at space-time point
   z[s] + E n, where z[s] is the world-line of the charge, E is
   a real number and n is a four-vector satisfying v[s].n = 0
   and n.n = 1 where v[s] is the four-velocity. The retarded
   proper-time corresponding to z[s] + E n is defined as s - y.
   Indices are represented by particular elements of a list so
   that, for example, a[1] and a[2] are the same four-vector
   but with differing indices. Any contractions (e.g a[1] a[1])
   are rewritten without their indices (a^2). *)

ps/: ps[(a_)[n_], m_] := ps[a[n],m] = Series[a[n][s - y], {y, 0, m}]
(* Expands any four-vector as a power series in y, about s. *)

k/: k[p_] := k[p] = E*n[p][s] + z[p][s] - ps[z[p], 4]
(* The vector joining z[s-y] and z[s] + E n. *)

vr/: vr[p_] := vr[p] = ps[Derivative[1][z[p]], 4]
(* Retarded velocity. *)

ar/: ar[p_] := ar[p] = ps[Derivative[1][Derivative[1][z[p]]], 3]
(* Retarded acceleration. *)

n[m_][s] := n[m]
z[n_][s] := z[n]
(z[n_])'[s] := v[n]
(z[n_])''[s] := a[n]
(z[n_])'''[s] := b[n]
(z[n_])''''[s] := c[n]
(z[n_])'''''[s] := d[n]
Unprotect[Power]
Unprotect[Times]

n[m_]^2 := 1
n[m_] v[m_] := 0
a[m_] n[m_] := an
b[m_] n[m_] := bn
c[m_] n[m_] := cn
d[m_] n[m_] := dn

v[n_]^2 := -1
a[n_] v[n_] := 0
b[n_] v[n_] := -a^2
c[n_] v[n_] := cv
d[n_] v[n_] := dv

a[n_]^2 := a^2
a[n_] b[n_] := ab
a[n_] c[n_] := ac
a[n_] d[n_] := ad

b[n_]^2 := b^2
b[n_] c[n_] := bc
b[n_] d[n_] := bd

c[n_]^2 := c^2
c[n_] d[n_] := cd

d[n_]^2 := d^2

yw[k_] := yw[k] = ExpandAll[k/. y -> w]

g[n_,m_] a_[m_] := a[n]
a_[m_] g[n_,m_] := a[n]

x := x = Series[(1 + E an)^(-1/2), {E, 0, 2}]
Protect[Power]
Protect[Times]

w = ExpandAll[E (1 + E^2((1/6) bn - (1/24) a^2)) x]
(* w is y in terms of E. *)

p  = - ExpandAll[yw[k[1]] yw[vr[1]]]
p1 = ExpandAll[p^-1]
p2 = ExpandAll[p1^2]
p3 = ExpandAll[p1^3]

kar = ExpandAll[yw[k[1]] yw[ar[1]]]

FI[n_, m_] := FI[n,m] =
ExpandAll[p3 (yw[k[m]] yw[vr[n]] - yw[k[n]] yw[vr[m]])]
(* Near field. *)

FII[n_, m_] := FII[n,m] =
ExpandAll[ExpandAll[p3 kar (yw[k[m]] yw[vr[n]] - yw[k[n]] yw[vr[m]])]
+ ExpandAll[p2 (yw[k[m]] yw[ar[n]] - yw[k[n]] yw[ar[m]])]]
(* Far field. *)

F[n_, m_] := F[n,m] = FI[n,m] + FII[n,m]
(* Total Field *)

Fsquared  = ExpandAll[F[1,2]^2]
FIsquared = ExpandAll[FI[1,2]^2]
FIIsquared = ExpandAll[FII[1,2]^2]
FIIIsquared = ExpandAll[2 FI[1,2] FII[1,2]]

G[n_]  := G[n]  = ExpandAll[F[n, n+1]  yw[v[n+1]]]
GI[n_] := GI[n] = ExpandAll[FI[n, n+1] yw[v[n+1]]]
GII[n_] := GII[n] = ExpandAll[FII[n, n+1] yw[v[n+1]]]

T[n_,m_] := T[n,m] = -ExpandAll[ (1/4)  ExpandAll[F[n,n+3]  F[n+3,m]]
+ (1/16) Fsquared g[n,m]]
(* Energy-momentum tensor. *)

TI[n_,m_] := TI[n,m] = -ExpandAll[(+1/4) ExpandAll[FI[n,n+3] FI[n+3,m]]
+ (1/16) FIsquared g[n,m]]
TII[n_,m_] := TII[n,m] =
-ExpandAll[(+1/4) Collect[Expand[Normal[FII[n,n+3]] Normal[FII[n+3,m]]], E]
+ (1/16) FIIsquared g[n,m]]
TIII[n_,m_] := TIII[n,m] =
-ExpandAll[(+1/4) (Collect[Expand[Normal[FI[n,n+3]]  Normal[FII[n+3,m]]], E]
+ Collect[Expand[Normal[FII[n,n+3]] Normal[FI[n+3,m]]], E])
+ (1/16) FIIIsquared g[n,m]]
TS[n_,m_] := TS[n,m] = ExpandAll[T[n,m] - TII[n,m]]
(* TS = Bound energy-momentum tensor. *)
(* TII = emitted energy-momentum tensor. *)

Tr[m_]    := Tr[m]    = ExpandAll[T[m,m+1]    n[m+1] (1 + E an) E^2]
TIr[m_]   := TIr[m]   = ExpandAll[TI[m,m+1]   n[m+1] (1 + E an) E^2]
TIIr[m_]  := TIIr[m]  = ExpandAll[TII[m,m+1]  n[m+1] (1 + E an) E^2]
TIIIr[m_] := TIIIr[m] = ExpandAll[TIII[m,m+1] n[m+1] (1 + E an) E^2]
TSr[m_] := TSr[m] = ExpandAll[Tr[m] - TIIr[m]]
(* Contractions of energy-momentum tensors. *)
\end{verbatim}
\endgroup

\chapter{SELF-FIELD THROUGH ANALYTIC CONTINUATION}
\label{app:B}
Recalling the discussion of \S\ref{sec:5.4}, we evaluate the retarded
field,
{
\setlength{\abovedisplayskip}{4pt plus 2pt minus 2pt}%
\setlength{\belowdisplayskip}{4pt plus 2pt minus 2pt}%
\begin{equation}
F^{\mu\nu}_{\mathrm{ret}}
= \frac{e}{\rho^{3}}\,v^{[\mu}r^{\nu]}
+ \frac{e}{\rho^{2}}\Bigl(a_{r}\,v^{[\mu}r^{\nu]} + a^{[\mu}r^{\nu]}\Bigr).
\label{eq:B.1}
\end{equation}
}
on the world-line of the charge by making the unphysical substitutions
\begin{align*}
x &= z(\tau),\\
z(\tau_{r}) &= z(\tau-\epsilon).
\end{align*}
The calculation is tedious although very straightforward; we simply expand
all retarded quantities in the right-hand side of (B.1) as power series in
$\epsilon$, and then when the calculation is complete we discard all terms of
order $\epsilon$ or greater.

The following \textit{Mathematica} program performs the calculation (note that
in the program $y \equiv \epsilon$, $k \equiv r$, $p \equiv \rho$, $kar \equiv \rho a_r$).

\begingroup
\begin{verbatim}
ps/: ps[(a_)[n_], m_] := ps[a[n],m] = Series[a[n][s - y], {y, 0, m}]

k/: k[p_] := k[p] = z[p][s] - ps[z[p], 4]
(* The vector z(s) - z(s-y) *)

vp/: vp[p_] := vp[p] = ps[Derivative[1][z[p]], 4]
(* v(s-y) *)

ap/: ap[p_] := ap[p] = ps[Derivative[1][Derivative[1][z[p]]],4]
(* a(s-y) *)

z[n_][s] := z[n]
(z[n_])'[s] := v[n]
(z[n_])''[s] := a[n]
(z[n_])'''[s] := b[n]
(z[n_])''''[s] := c[n]
(z[n_])'''''[s] := d[n]
Unprotect[Power]
Unprotect[Times]

v[n_]^2 := -1
a[n_] v[n_] := 0
b[n_] v[n_] := -a^2
c[n_] v[n_] := cv
d[n_] v[n_] := dv

a[n_]^2 := a^2
a[n_] b[n_] := ab
a[n_] c[n_] := ac
a[n_] d[n_] := ad

b[n_]^2 := b^2
b[n_] c[n_] := bc
b[n_] d[n_] := bd

c[n_]^2 := c^2
c[n_] d[n_] := cd

d[n_]^2 := d^2

Protect[Power]
Protect[Times]

p  = - ExpandAll[k[1] vp[1]]
p1 = ExpandAll[p^(-1)]
p2 = ExpandAll[p1^2]
p3 = ExpandAll[p1^3]

kar = ExpandAll[k[1] ap[1]]

FI[n_, m_] := FI[n,m] = ExpandAll[p3 (k[m] vp[n] - k[n] vp[m])]
(* Near Field *)

FII[n_, m_] := FII[n,m] = ExpandAll[ExpandAll[p3 kar (k[m] vp[n] - k[n] vp[m])]
+ ExpandAll[p2 (k[m] ap[n] - k[n] ap[m])]]
(* Far Field *)

F[n_, m_] := F[n,m] = ExpandAll[FI[n,m] + FII[n,m]]
(* Total Field *)

G[n_] := G[n] = ExpandAll[F[n, n+1] v[n+1]]
\end{verbatim}
\endgroup

From the above program we find
\begin{equation}
F^{\mu\nu}_{\mathrm{ret}}
= e\left[\lim_{\epsilon\to 0}\frac{a^{[\mu}v^{\nu]}}{2\epsilon}
+ \frac{2}{3}v^{[\mu}\dot{a}^{\nu]}\right].
\label{eq:B.2}
\end{equation}
The advanced field is clearly just obtained from the retarded field by changing
the sign of $\epsilon$ and the sign of the whole expression.
\begin{equation}
F^{\mu\nu}_{\mathrm{adv}}
= e\left[\lim_{\epsilon\to 0}\frac{a^{[\mu}v^{\nu]}}{2\epsilon}
- \frac{2}{3}v^{[\mu}\dot{a}^{\nu]}\right].
\label{eq:B.3}
\end{equation}

\clearpage
\begin{center}
  {\LARGE\bfseries ACKNOWLEDGEMENTS}
\end{center}
I would like to thank my supervisor, Dr.\ Peter Szekeres, for very
helpful discussions during the year. I am also very grateful to
Belinda Medlyn for her painstaking proofreading of the manuscript. All
errors are mine.

\addcontentsline{toc}{chapter}{Bibliography}
\vspace{-3cm}


\begin{thebibliography}{12}
\bibitem{Barut1964}
A. O. Barut. \textit{Electrodynamics and Classical Theory of Fields and Particles} (Macmillion \& co, 1964).

\bibitem{Rohrlich1965}
F. Rohrlich. \textit{Classical Charged Particles} (Addison-Wesley Publishing Co., 1965).

\bibitem{Dirac1938}
P. A. M. Dirac, Proc. Roy. Soc. (London) \textbf{A167}, 148 (1938).

\bibitem{Teitelboim1970}
C. Teitelboim, Phys. Rev. \textbf{D1}, 1572 (1970).

\bibitem{Teitelboim1971}
C. Teitelboim, Phys. Rev. \textbf{D4}, 345 (1971).

\bibitem{Barut1974}
A. O. Barut, Phys. Rev. \textbf{D10}, 3335 (1974).

\bibitem{BarutVillarroel1975a}
A. O. Barut, D. Villarroel, J. Phys. A. \textbf{8}, 156 (1975).

\bibitem{BarutVillarroel1975b}
A. O. Barut, D. Villarroel, J. Phys. A. \textbf{8}, 1537 (1975).

\bibitem{HawkingEllis73}
S. W. Hawking, R. E. Ellis, \textit{The Large Scale Structure of Space Time}.

\bibitem{LandauLifshitz80}
L. Landau, L. Lifschitz, \textit{The Classical Theory Of Fields}.

\bibitem{Sorg1978}
M. Sorg, Z. Naturforsch. \textbf{33a}, 619 (1978).

\bibitem{TeitelboimReviewUnknown}
A review paper by Teitelboim et al. the origin of which is unknown to me.
\end{thebibliography}
\end{document}